\documentclass[article,10pt]{IEEEtran} 

\usepackage{etex} 
\usepackage{amsmath,amsfonts,amssymb}
\usepackage{cite}
\usepackage{balance} 
\usepackage{multirow} 
\usepackage{pbox} 
\usepackage[caption=false]{subfig} 
\usepackage{mathrsfs} 
\usepackage{amsbsy} 
\usepackage{bm} 
\usepackage{mathtools} 
\usepackage{paralist} 
\usepackage{tabularx} 
\usepackage{dcolumn} 
\usepackage{anyfontsize} 
\usepackage{enumitem} 
\usepackage{amsthm} 
\usepackage{stfloats} 
\usepackage{caption} 

\usepackage[utf8]{inputenc}
\usepackage{cancel}
\usepackage{tabto}
\usepackage{bm}
\usepackage{float}
\usepackage{rotating} 
\usepackage{array}               
\usepackage{pdfpages} 
\usepackage{subfig}
\usepackage{floatflt}   
\usepackage{subfiles}
\usepackage{colortbl}
\usepackage[export]{adjustbox}

\usepackage{grffile}

\definecolor{matlabBlue}{rgb}{0.00000,0.44700,0.74100}%
\definecolor{matlabOrange}{rgb}{0.85000,0.32500,0.09800}%
\definecolor{matlabYellow}{rgb}{0.92900,0.69400,0.12500}%
\definecolor{matlabLila}{rgb}{0.49400,0.18400,0.55600}%
\definecolor{matlabGreen}{rgb}{0.46600,0.67400,0.18800}%
\DeclareMathOperator*{\argmax}{argmax}

\newcommand{\rmv}{\hspace*{-.3mm}}
\newcommand{\iist}{\hspace*{1mm}}

\providecommand{\keywords}[1]{\textbf{\textit{Keywords---}} #1}

\begin{document}
\frenchspacing

\title{RSS-based Cooperative Localization and Orientation Estimation Exploiting Directive Antenna Patterns}
\author{Lukas Wielandner$^{1,2}$, Erik Leitinger$^{1}$, Klaus Witrisal$^{1,2}$
\thanks{This work was supported in part by the Christian Doppler Research Association, the
Austrian Federal Ministry for Digital and Economics Affairs and the National
Foundation for Research, Technology and Development, by SES imagotag GmbH, and by the Austrian Science Fund (FWF) under grant J 4027.}
\\
\small{{$^1$Graz University of Technology}}\\
\small{{$^2$Christian Doppler Laboratory for Location-aware Electronic Systems}}}
\maketitle

\begin{abstract}
In this paper, we propose a factor-graph-based cooperative positioning algorithm that uses RSS radio measurements and accounts for the directivity of the antennas. This is achieved by modeling the directivity with a parametric antenna pattern and jointly estimating position and orientation of the agents. We propose two different approaches whereas the first one uses a continuous representation of the orientation state and the second one a discrete representation. We validate our proposed methods with simulations and measurements in a static sensor network with more than 900 agents in an indoor environment and show that the positioning accuracy can be improved significantly by considering the influence of orientations.
\end{abstract}


\section{Introduction}
For the Internet-of-things (IoT), location awareness is a key aspect for various applications \cite{DiTaranto2014SPM,win2018IoT}. One example are large, static networks, which are gaining importance in logistics, industry and retailing \cite{da2014internet,ngai2008rfid}. They are used to label and identify goods and can be utilized to find packages, tools or products of interest. Conventionally, the positions of these objects have to be added to a database manually, which is highly inefficient. This problem can be avoided by using a technology that allows to infer the positions of the objects using measurements to base stations with known positions (anchors) and measurements in-between the objects (agents) in the network, employing cooperative localization algorithms \cite{patwari2005locating,CorWarJurHuValMoo:ProcIEEE2010,wymeersch2009cooperative,meyer2015distributed_Loc,WinConMC2011,win2018IoT}. Cooperative localization leads to an improvement of positioning accuracy while preventing the use of a high-density anchor deployment as needed for non-cooperative localization \cite{alsindi2007cooperative,ShenTIT2010part2,patwari2005locating}. Many localization systems  in IoT scenarios use time-of-arrival (ToA) \cite{meyer2015distributed_Loc,MeyerJIT2018}, angle-of-arrival (AoA) \cite{fascista2017angle,wu2020cooperative} or received signal strength (RSS) measurements \cite{BestRSSMeasurements_IEEE_Survey2016,LongRangeIoT_IWC_2016}. ToA-based and AoA-based methods have often a higher localization accuracy compared to RSS-based methods \cite{alam2013cooperative}. However, they require exact synchronization \cite{MeyerJIT2018, EtzlingerTSP2017} or coherent processing \cite{HanTIT2016,ShenTIT2010part2}, which can only be achieved with expensive equipment. In spite of the high localization performance, for real applications, cost efficiency is one of the key aspects for a technology to be widely used. Therefore RSS measurements are an obvious choice for low-cost sensors with limited power supply. Based on the RSS, one can get an estimate of the distance between the sensors, which can be used to infer the positions of the agents \cite{Slavisa_TVT_2015}. However, in indoor environments characterized by severe multipath fading, only a low positioning accuracy is achieved, with a large number of outliers, due to the high uncertainty of the measurement data. Furthermore, obstructed line-of-sight (OLOS) results in large fluctuations of the path-loss exponent. The RSS model is very sensitive to these fluctuations leading to large ranging errors. To overcome these impairments, much effort is put into the development of robust RSS-based localization methods. 
\subsection{State of the Art and Related Work}
An important aspect of RSS-based localization is the difference between non-model-based and model-based approaches, which have various advantages and disadvantages. Non-model-based approaches are for example fingerprinting \cite{RSSI_fingerprinting_2017,SavicVTC2015} and machine learning \cite{PrasadHossainBhargavaTWC2018,VieiraPIMRIC2017,BurghalArxiv2020comprehensive}. Fingerprinting consist of an offline training phase and an online localization phase. In the training phase, measurements are acquired at known positions and stored in a database. Localization is performed by comparing measurements to labeled measurements stored in the database. The main disadvantages of non-model-based approaches are the requirement of a large number of labeled measurements, collected in time-consuming measurement campaigns, and that the database can only be used for the according environment, i.e., the trained model based on the database does not generalize to other environments \cite{HuangMolischHeWangTangAiZhongTWC2020}. The same is true for machine learning based approaches. However, their main advantages are that they generalize to arbitrary models and that their feed-forward execution has rather low computational complexity, in comparison to fingerprinting and also to model-based approaches.

Model-based approaches do not require training data and they are more or less independent of the surrounding environment as long as the chosen model agrees well with the measured data.     
There exists a variety of different model-based approaches for RSS-based localization. In \cite{LiAsilomar2020_RSS, li2020target} an algorithm has been introduced that utilizes an antenna array with known orientation and antenna pattern to measure the RSS differences, allowing to exploit the AoA non coherently. Another example are ranging methods \cite{Slavisa_TVT_2015,yang2009indoor}, which use a path-loss model to estimate the distance in-between the involved nodes. Methods that include the possibility to receive OLOS measurements in the modelling process can counteract the degradation due to multipath and path-loss variations \cite{wang2011new,sun2019robust}. A different approach for achieving a more robust RSS-based localization takes additional effects like the radiation pattern of the antennas into account \cite{zuo2019rss, RSSD_AOA_ISOC_2012, Jia_RSSD_2018}.

For large sensor networks with multiple measurements in-between the sensors, cooperative localization is a common approach for estimating the states of static and mobile agents. A requirement for cooperative localization is that agents can interact with each other and exchange information about their state. This can be achieved in a distributed or centralized manner \cite{wymeersch2009cooperative,meyer2015distributed_Loc}. Theoretical performance limits for cooperative localization can be found in \cite{ShenJSAC2012,ShenTIT2010part2,WinProc2018TheoreticalFoundCoopLoc} based on the equivalent Fisher information. 
When dealing with large networks, full cooperation is sometimes not wanted, due to resource limitations with regard to energy consumption, measurement time, and computational power. In these cases, selecting highly informative cooperating partners can be essential since at some point increasing the set of cooperating partners does not increase the positioning performance any more \cite{win2018network,ShenTIT2010part2,wielandner2020information}. 
For range-based CL, there exists a variety of different approaches like convex optimization \cite{Slavisa_TVT_2015} or a fully Bayesian treatment of the problem as in \cite{ihler2005nonparametric,meyer2015distributed_Loc,Naseri2019,Etzlinger2017_CoSLAS,Wang2020_NLOSmitigation,Etzlinger2020_WSN}. In \cite{Wu2020_postLin}, the authors use only angular information based on coherent antenna array processing for cooperative localization. To estimate the states of the agents in a Bayesian framework, factorization of the joint agent states probability density function (PDF) is essential to reduce the complexity. A common approach for visualizing the factorization is by using a factor graph \cite{KschischangTIT2001, Loeliger2004SPM}. Based on the factor graph representation of the problem, one can use message passing algorithms like parametric or nonparametric belief propagation to iteratively estimate the states of the agents \cite{meyer2013sigma,Wu2020_postLin,ihler2005nonparametric,meyer2015distributed_Loc}. Cooperative localization algorithms that use RSS measurements and are based on message passing are given in \cite{Jin2016_RSS_BP,Jin2020_RSS_CL}.
Taking into account additional information like the radiation pattern of the antennas in the modeling process and in the message passing scheme, and therefore extending a range-based cooperative localization algorithm, is a logical step in the search for robust RSS-based localization.

\subsection{Contributions}
In this paper, we develop a Bayesian message passing algorithm for cooperative localization based on a factor graph that includes the directivity of the antennas to exploit AoA information in a non-coherent way. We propose two different methods for estimating position and orientation of the agents in a Bayesian framework. The first method estimates the position and the orientation in a continuous parameter space, using a particle-based representation of the states \cite{DoucetSPM2005,arulampalam2002tutorial}. The second one uses a discrete parameter space for the orientations and continuous parameter space for the positions. We develop a factor graph representation for both methods. The estimation of positions and orientations is based on the marginal posterior PDFs of the states, which are calculated using belief propagation. We validate our proposed methods with simulations and real RSS measurements for more than 900 agents in an indoor environment and show that the positioning accuracy can be improved significantly by considering the influence of orientations. To the best of our knowledge, the use of angle information by utilizing directive antennas has not been investigated for cooperative localization. The advantage of this approach is that we have a more realistic model for the RSS measurements. However the disadvantage compared to the standard path-loss model is that the antenna pattern has to be known or estimated and the orientation of the nodes has to be estimated as well for the antenna pattern to be useful. The key contributions of this paper are as follows.
\begin{itemize}
\item We develop a Bayesian message passing algorithm based on a factor graph for joint cooperative position and orientation estimation.
\item We compare the inference models for a continuous and a discrete representation of the orientation state.
\item We use a parametric antenna pattern and estimate model order and parameters based on the measurements.
\item We give a comprehensive analysis of the presented algorithms using synthetic and measured data.
\end{itemize}
We would like to highlight that the focus of this paper is on static networks, however, it is straightforward to extend the proposed algorithms to a dynamic scenario as shown in \cite{meyer2015distributed_Loc}. We organize the remainder of this paper as follows. In Section~\ref{sec:SystemModel}, we state the system model and introduce the measurement model. The difference between the continuous and the discrete representations of the orientation state as well as the likelihood function is explained in Section~\ref{sec:infM}. In Section~\ref{sec:BP} we give an introduction on belief propagation and develop the proposed algorithms. The results of numerical experiments and measurements are reported in Sections~\ref{sec:Results} and \ref{sec:Meas_Results}, where we also explain how we determine the antenna pattern. Section~\ref{sec:Conclusion} concludes the paper.
\section{System Model}
\label{sec:SystemModel}

We consider a set of static cooperating agents $\mathcal{C}$ with unknown positions and orientations ($|\mathcal{C}|$ is the number of agents) and a set of anchors $\mathcal{A}$ with known positions and orientations ($|\mathcal{A}|$ is the number of anchors). For this we define two types of measurements: (i) measurements between anchors and agents $\bar{z}_{a,i}$ with $a \in \mathcal{A}$ and $i \in \mathcal{C}$ and (ii) measurements in-between agents $\underline{z}_{i,j}$ with $i \in \mathcal{C}$ and $j \in \mathcal{D}_i$ with $\mathcal{D}_i \subseteq \mathcal{C}$ being the set of agents that cooperate with agent $i$. The stacked vector of all anchor measurements is written as $\bar{\mathbf{z}} = [\bar{z}_{1,1} \iist \bar{z}_{1,2}  \iist \dots \iist \bar{z}_{1,|\mathcal{C}|} \iist  \bar{z}_{2,1} \iist \dots \iist \bar{z}_{|\mathcal{A}|,|\mathcal{C}|}]^T$ which can also be written as $\bar{\mathbf{z}} = [\bar{z}_{a,i}]_{a \in \mathcal{A}, i \in \mathcal{C}}$. The stacked vector of all measurements in-between agents is given as $\underline{\mathbf{z}} = [\underline{z}_{i,j}]_{i \in \mathcal{C},j \in \mathcal{D}_i}$. Each node (agent or anchor) has a fixed position and orientation and a known antenna pattern (which does not vary with time). We assume that the antenna patterns of the nodes are all identical.

\subsection{Agent State Model}
The state of the $i$th agent, denoted $\boldsymbol{\theta}_i \in \mathbb{R}^{3 \times 1}$ is defined as $\boldsymbol{\theta}_i = [\mathbf{x}_i \iist\varphi_i]^\text{T} $ and consists of its position ${\mathbf{x}_i = [x_i \iist y_i]}$ and the orientation $\varphi_i$. Note that the anchor state is defined in the same way, however, it is assumed to be known.
The vectors $\boldsymbol{\theta}$ and $\mathbf{z}$ denote the stacked vectors of all agent and anchor states and all measurements (between agents and anchors and agents and agents) respectively, where ${\boldsymbol{\theta} = [\boldsymbol{\theta}_1 \iist \dots \iist \boldsymbol{\theta}_{|\mathcal{C}|}, \boldsymbol{\theta}_{|\mathcal{C}|+1}\iist \dots \iist \boldsymbol{\theta}_{|\mathcal{C}|+|\mathcal{A}|}]^T}$ and $\mathbf{z} = [\bar{\mathbf{z}}^T \iist \underline{\mathbf{z}}^T]^T$.
The joint posterior PDF is given as
\begin{align}
f(\boldsymbol{\theta}|\mathbf{z}) &\propto f(\mathbf{z}|\boldsymbol{\theta}) f(\boldsymbol{\theta}).
\label{eq:contJointPost}
\end{align}
with the joint likelihood $f(\mathbf{z}|\boldsymbol{\theta})$ and the prior of all agent and anchor states $f(\boldsymbol{\theta})$.

\subsection{Measurement Model}
\label{sec:measModel}
\begin{figure}[t]
\centering
\includegraphics[scale=1]{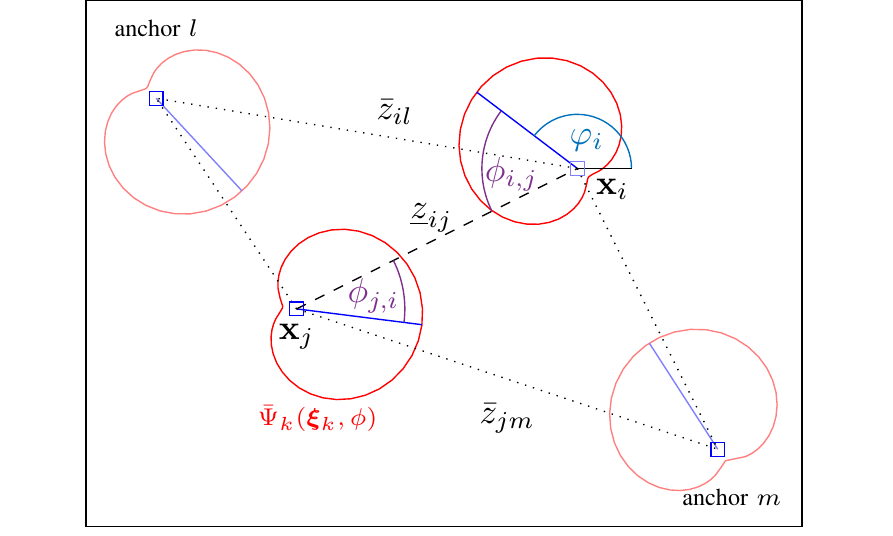}
\caption{This figure shows two agents and two anchors which are arbitrary orientated to one another. The centers of the nodes and the orientations are indicated by blue squares and solid blue lines, respectively. The antenna pattern $\bar{\Psi}_k(\boldsymbol{\xi}_k,\phi)$ is given in red. The dashed line indicates the measurement between two agents whereas the dotted line corresponds to measurements between anchors and agents.}
	\label{fig:AP}
\end{figure}

We use RSS measurements to infer position and orientation of the agents. For that purpose, we model the RSS as a combination of the distance-depending path-loss and the influence of the orientation of the agents which have performed the measurements. The measurement between two nodes $i$ and $j$ is conditioned on the measurement model $M$. In general the measurement model is a discrete random variable, taking values of model indices $k=\{1,2,\dots,N_{\mathcal{M}}\}$. The members of the set are indicating different models with $N_{\mathcal{M}}$ being the number of models. In this section, we assume that the model $M\!=\!k$ as well as the model parameters are known. Different models are discussed in Section~\ref{sec:detAP}. The measurement conditioned on the model index $k$ has the form of 
\begin{align}
z_{i,j}|k &\iist = L_k(P_k,n_k,d_{i,j}) + \Psi_k(\boldsymbol{\xi}_k,\phi_{i,j},\phi_{j,i}) + n_{i,j}
\label{eq:measModelAP}
\end{align}
with the distance-depending path-loss $L_k$, the resulting change of the RSS in dependency of the orientation of the agents $\Psi_k$ and a noise term that is assumed to be Gaussian distributed as $n_{i,j}\! \sim \!\mathcal{N}(0,\sigma_k^2)$. We make the common assumption that interfering multipath components, which lead to fading, as well as shadowing are only described by the noise term \cite[Section 3.9.2]{rappaport1996wireless}. 
The distance-depending path-loss in dB is given as
\begin{equation}
L_k(P_k,n_k,d_{i,j}) = 10\text{log} \left(10^{P_k/10} \left(\frac{d_{i,j}}{d_0}\right)^{-n_k}\right)
\label{eq:pathloss}
\end{equation}
where $n_k$ and $P_k$ are the path-loss exponent and the reference path-loss at distance $d_0$ of model $k$ respectively \cite{patwari2005locating}. The distance between the nodes is given as ${d_{i,j} = \|\mathbf{x}_j^\text{T} - \mathbf{x}_i^\text{T}\|}$ where $\|\cdot\|$ indicates the Euclidean norm. The resulting change of the measured signal strength, which depends only on the orientation of two agents, can be written as the sum of the two antenna patterns evaluated at the corresponding angles $\phi_{i,j}$ and $\phi_{j,i}$ as
\begin{align}
\Psi_k(\boldsymbol{\xi}_k,\phi_{i,j},\phi_{j,i}) = &  \bar{\Psi}_k(\boldsymbol{\xi}_k,\phi_{i,j}) + \bar{\Psi}_k(\boldsymbol{\xi}_k,\phi_{j,i}) 
\label{eq:2wayAP}
\end{align}
where $\phi_{i,j}$ is the angle between agent $i$ and agent $j$ with respect to the orientation $\varphi_i$ of agent $i$ and $\phi_{j,i}$ is the angle in-between agent $j$ and agent $i$ with respect to the orientation $\varphi_j$ of agent $j$. The generic model of an antenna pattern is indicated as $\bar{\Psi}_k(\boldsymbol{\xi}_k,\phi)$. The graphical representation of a measurement in-between two agents can be seen in Fig.~\ref{fig:AP}. 
The model parameters of the antenna pattern $\bar{\Psi}_k$ are stacked in the vector ${\boldsymbol{\xi}_k = [\xi_{k,1},\dots,\xi_{k,N_k}]^T}$ with $N_k$ being the number of parameters of the $k$th antenna pattern.
We define the angle between two nodes as ${\phi_{i,j}=\text{atan2}(y_j - y_i, x_j - x_i) - \varphi_i}$. The used antenna patterns are introduced in Sections~\ref{sec:Results} and~\ref{sec:Meas_Results}.

\section{Inference models}
\label{sec:infM}
In this section we introduce two different inference models to estimate position and orientation of the agents based on their marginal PDFs. The first one is referred to as  ``continuous inference model'' since it uses a continuous representation for the agent positions and orientations. The second inference model is called ``discrete inference model'', where the only difference to the continuous inference model is that the orientation estimation is performed with a set of discrete orientations instead of a continuous parameter space.

Since we are estimating position and orientation based on the marginal posterior PDF of each agent state, we define at this point the \textit{not-integrate} and \textit{not-sum} notations as mentioned in \cite{KschischangTIT2001}. Instead of indicating which variables are marginalized over, we define the ones over which we do not sum or integrate. For example if $h(\cdot)$ is a function that depends on the continuous variables $\boldsymbol{\theta}_1$, $\boldsymbol{\theta}_2$ and $\boldsymbol{\theta}_3$ and $g(\cdot)$ is a function that depend on the discrete variables $\Phi_1$, $\Phi_2$ and $\Phi_3$, the \textit{not-integrate} and \textit{not-sum} for $\boldsymbol{\theta}_3$ and $\Phi_3$ are defined as
\begin{align}
\int h(\boldsymbol{\theta}_1,\boldsymbol{\theta}_2,\boldsymbol{\theta}_3) d\boldsymbol{\theta}_{\sim 3} & \coloneqq \iint h(\boldsymbol{\theta}_1,\boldsymbol{\theta}_2,\boldsymbol{\theta}_3) d\boldsymbol{\theta}_1 d\boldsymbol{\theta}_2 \label{eq:notInt}\\
\sum_{\Phi_{\sim 3}}  g(\Phi_1,\Phi_2,\Phi_3) & \coloneqq \sum_{\Phi_1 \in \mathcal{O}} \sum_{\Phi_2 \in \mathcal{O}} g(\Phi_1,\Phi_2,\Phi_3) \label{eq:notSum}
\end{align}
where $\mathcal{O}$ is the domain of $\Phi_i$.

\subsection{Continuous inference model}
\label{sec:cont_infM}
For the continuous inference model, we factorize the joint posterior PDF in \eqref{eq:contJointPost} as
\begin{align}
f(\boldsymbol{\theta}|\mathbf{z}) &\propto  \prod_{i \in \mathcal{C}} f(\boldsymbol{\theta}_i) \prod_{a \in \mathcal{A}}  f(\bar{z}_{a,i}|\boldsymbol{\theta}_{i};\boldsymbol{\theta}_{a})\prod_{j \in \mathcal{D}_i}  f(\underline{z}_{i,j}|\boldsymbol{\theta}_{i},\boldsymbol{\theta}_{j})
\label{eq:joint_post}
\end{align}  
since the measurements between nodes are independent of each other. The factorization consists of the prior PDF $f(\boldsymbol{\theta}_i)$ of agent state $i$, the likelihood function $f(\bar{z}_{a,i}|\boldsymbol{\theta}_{i};\boldsymbol{\theta}_{a})$ which is a function of $i$th agent state and parametrized by the $a$th anchor state (which is known) and the ``cooperation'' likelihood function $f(\underline{z}_{i,j}|\boldsymbol{\theta}_{i},\boldsymbol{\theta}_{j})$ which is a function of the $i$th agent state and $j$th agent state with $i\neq j$. Since the states of the anchors  are exactly known and independent of all other states, we can rewrite \eqref{eq:joint_post} as
\begin{equation}
f(\boldsymbol{\theta}|\mathbf{z}) \propto  \prod_{i \in \mathcal{C}} f(\boldsymbol{\theta}_i|\bar{\mathbf{z}}) \prod_{j \in \mathcal{D}_i}  f(\underline{z}_{i,j}|\boldsymbol{\theta}_{i},\boldsymbol{\theta}_{j})
\end{equation}
where $f(\boldsymbol{\theta}_i|\bar{\mathbf{z}})$ is the marginal posterior PDF of the $i$th agent state given all anchor measurements. In the Section~\ref{sec:BP} the factor graph representation of this equation will be described.  
With the help of the \textit{not-integrate} notation in \eqref{eq:notInt}, the marginal posterior PDF of state $\boldsymbol{\theta}_i$ is given by
\begin{equation}
f(\boldsymbol{\theta}_i|\mathbf{z}) = \int f(\boldsymbol{\theta}|\mathbf{z}) d\boldsymbol{\theta}_{\sim i}
\label{eq:marginal_post_cont}
\end{equation}
where $\int d\boldsymbol{\theta}_{\sim i}$ denotes the marginalization of all states $\boldsymbol{\theta}$ except $\boldsymbol{\theta}_{i}$.

\subsection{Discrete inference model}
\label{sec:disc_infM}
For the discrete inference model, we have a continuous state of the $i$th agent, denoted $\mathbf{x}_i \in \mathbb{R}^{2 \times 1}$, which consists of its position ${\mathbf{x}_i = [x_i \iist y_i]^T}$ and, independent of it, a discrete random variable $\Phi_i$, taking values from the finite set ${\mathcal{O}=\{\tilde{\varphi}_1, \tilde{\varphi}_2, \dots ,\tilde{\varphi}_{N_\mathcal{O}}\}}$ with the number of discrete orientations given as $N_\mathcal{O}$. The probabilities for the discrete orientations of agent $i$ are indicated by a probability mass function (PMF) as $p(\Phi_i)$. The joint posterior PDF is given as
\begin{align}
f(\mathbf{x}, \boldsymbol{\Phi}|\mathbf{z}) & \propto f(\mathbf{z}|\mathbf{x}, \boldsymbol{\Phi}) f(\mathbf{x}) p(\boldsymbol{\Phi})
\end{align}
with $\mathbf{x} = [\mathbf{x}_i]_{i \in \mathcal{C}}$ and $\boldsymbol{\Phi} = [\Phi_i]_{i \in \mathcal{C}}$ being the stacked vector of all position states and discrete orientation states, respectively. The joint posterior PDF factorizes into
\begin{align}
f(\mathbf{x}, \boldsymbol{\Phi}|\mathbf{z}) \propto & \prod_{i \in \mathcal{C}} f(\mathbf{x}_i) p(\Phi_i) \nonumber \\
& \times \prod_{a \in \mathcal{A}}  f({\bar{z}}_{a,i}|\mathbf{x}_i, \Phi_i;\mathbf{x}_a, \Phi_a) \nonumber \\
& \times \prod_{j \in \mathcal{D}_i}  f(\underline{z}_{i,j}|\mathbf{x}_i, \Phi_i,\mathbf{x}_j, \Phi_j)
\label{eq:jointPostFactorDiscrete}
\end{align}  
which consists of the prior PDF of the position states, the prior PMF of the orientation states as well as the likelihood functions regarding measurements to anchors and to cooperating agents. We want to mention that for the discrete inference model, we evaluate the likelihood for every possible combination of orientations between agent $i$ and agent $j$ out of the set of discrete orientations $\mathcal{O}$.
Given the definitions in \eqref{eq:notInt} and \eqref{eq:notSum}, the marginal posterior PDF of agent $i$ can then be written as
\begin{equation}
f(\mathbf{x}_i, \Phi_i|\mathbf{z}) = \sum_{\Phi_{\sim  i}} \int f(\mathbf{x}, \boldsymbol{\Phi}|\mathbf{z}) d\mathbf{x}_{\sim i}
\label{eq:margPostFactorDiscrete}
\end{equation}
where $\sum_{\Phi_{\sim  i}}$ and $\int  d\mathbf{x}_{\sim i}$ denote the marginalization of all states except of $\Phi_{i}$ and $\mathbf{x}_{i}$, respectively.

\subsection{Likelihood} 
We define the model parameter vector for measurement model $k$ as ${\boldsymbol{\vartheta}_k = [P_k,n_k,\boldsymbol{\xi}_k^T,\sigma_k]^T}$ that consists of the reference path-loss $P_k$ at distance $d_0$, the path-loss exponent $n_k$, the model parameters of the antenna pattern $\boldsymbol{\xi}_k$ and the shadowing standard deviation $\sigma_k$.
For a Gaussian noise model given in \eqref{eq:measModelAP}, the likelihood function can be written in the form of
\begin{equation}
f(z_{i,j}|\boldsymbol{\theta}_{i},\boldsymbol{\theta}_{j},\boldsymbol{\vartheta}_k,M\!\!=\!\! k) = \frac{1}{\sqrt{2\pi \sigma_k^2}} \text{exp}\left(\frac{1}{2\sigma_k^2} (z_{i,j} - \tilde{z}_{i,j})^2  \right)
\label{eq:lhf}
\end{equation}
with
\begin{equation}
\tilde{z}_{i,j} = L_k(P_k,n_k,d_{i,j}) + \Psi_k(\boldsymbol{\xi}_k,{\phi}_{i,j} ,{\phi}_{j,i})
\label{eq:lhf_param}
\end{equation}
and $\boldsymbol{\theta}_i = [\mathbf{x}_i \iist\varphi_i]^\text{T} $. Note that model $k$ as well as the parameters $\boldsymbol{\vartheta}_k$ are assumed to be known which results in a likelihood function that depends only on the states of the nodes $f(z_{i,j}|\boldsymbol{\theta}_{i},\boldsymbol{\theta}_{j})$. The different models as well as the estimation of the parameters are discussed in Section~\ref{sec:detAP}. 
\section{Message Passing Algorithms}
\label{sec:BP}
In a Bayesian framework, we estimate position and orientation of each agent based on the marginal posterior PDFs. Since a straightforward computation of the marginal posterior PDFs $f(\boldsymbol{\theta}_i|\mathbf{z})$ and $f(\mathbf{x}_i, \Phi_i|\mathbf{z})$ in \eqref{eq:marginal_post_cont} and \eqref{eq:margPostFactorDiscrete} is infeasible, we perform message passing by means of the sum-product-algorithm rules on the factor graph that represents our statistical model. This so called ``belief propagation'' yields approximations (``beliefs'') of the marginal posterior PDF in an efficient way \cite{KschischangTIT2001, Loeliger2004SPM}. It gives the exact marginal PDF for a tree like graph but provides only an approximate marginalization if the underlying factor graph has cycles \cite{KschischangTIT2001}. In this case, the belief propagation scheme becomes iterative and there exist different orders in which the messages can be calculated. We have chosen that in each iteration, the beliefs of all agents $i\in \mathcal{C}$ are updated in parallel. In the following section, we derive the belief propagation message passing scheme for the continuous and the discrete inference models based on the factor graphs in Fig.~\ref{fig:FG}.
\subsection{Belief propagation for continuous model}
\label{sec:BP_cont}
\begin{figure}[t]
\subfloat[Continuous inference model ]{
{\includegraphics[scale=0.91]{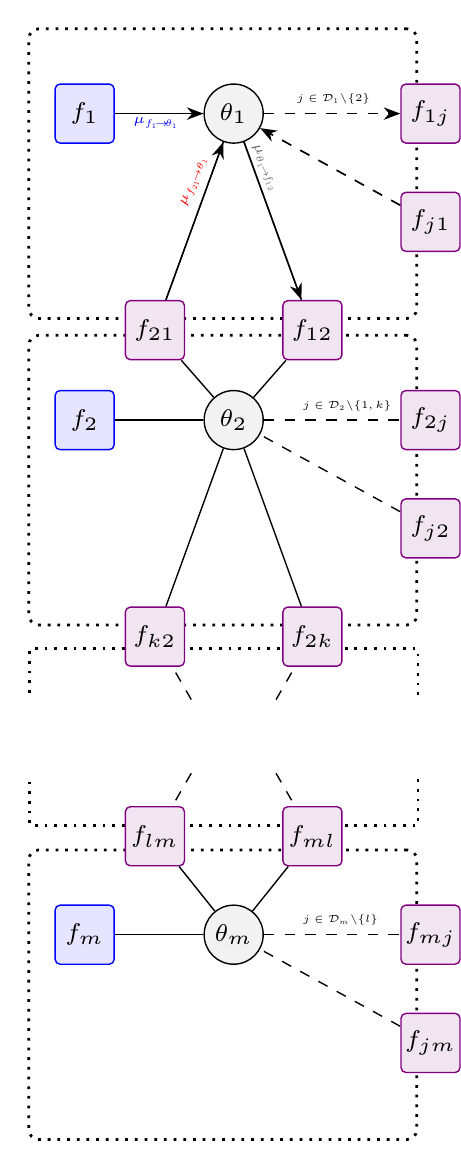}\label{fig:FG_cont}}}
\subfloat[Discrete inference model]{
{\includegraphics[scale=0.91]{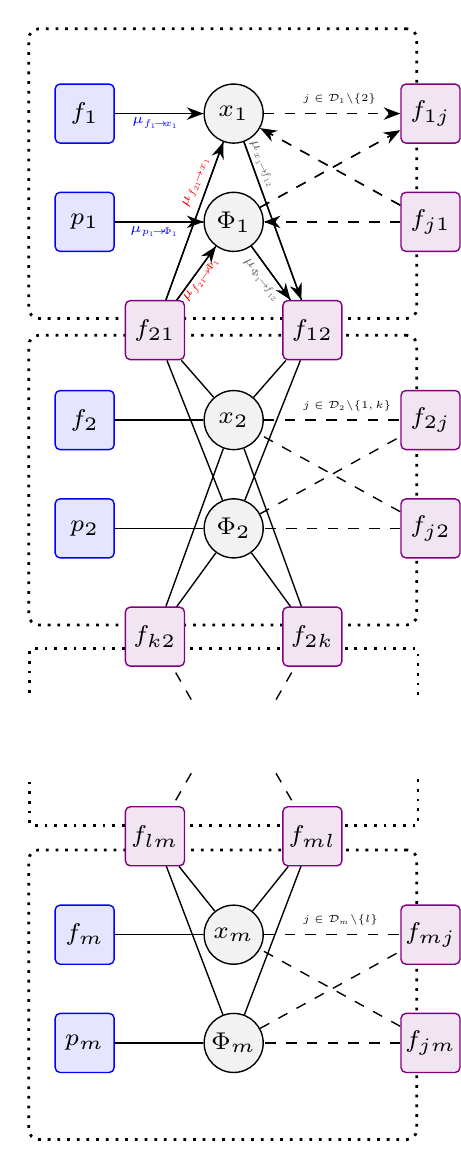}\label{fig:FG_disc}}}
\caption{This figure shows the factor graph for the two proposed cooperative localization algorithms. On the left side, we see the factor graph for a complete continuous representation of the agent state $\theta$. On the right hand side we see the factor graph for a mixture of continuous and discrete representation of the agent state. The factors $f_i$ and $p_i$ represent prior information about the $i$th agents state. In our case, prior information is given by geometrical boundaries, possible orientations and information from measurements to anchors. The notation $\mathcal{D}_m \backslash \{l\}$ means all members of  $\mathcal{D}_m$ except $l$.}
\vspace{-3mm}
\label{fig:FG}
\end{figure}

Based on the factor graph for the continuous inference model given in Fig.~\ref{fig:FG_cont}, we define the message passing scheme to approximate the marginal posterior PDFs. For a better readability, we use the following short hand notation: The factor $f_{ij}~\triangleq~f(\underline{z}_{i,j}|\boldsymbol{\theta}_{i},\boldsymbol{\theta}_{j})$ represents the likelihood function with respect to the involved agents $i$ and $j$ whereas $f_{i}~\triangleq~f(\boldsymbol{\theta}_{i})\prod_{a \in \mathcal{A}}  f(\bar{z}_{a,i}|\boldsymbol{\theta}_{i};\boldsymbol{\theta}_{a})$ represent prior information of the $i$th agent state and information from the anchor measurements. Additionally, we define $\mu_{\theta_i \rightarrow f_{ij}}$ as the message from variable node $\theta_i$ to factor node $f_{ij}$, $\mu_{f_{ji}  \rightarrow \theta_i}$ as the message from factor node $f_{ji}$ to variable node $\theta_i$ and $\mu_{f_i \rightarrow \theta_i}$ as the message from factor node $f_{i}$ to variable node $\theta_i$. The belief of the state of agent $i$ is written as $m(\boldsymbol{\theta}_i)$ and is calculated by multiplying all the incoming messages at variable node $\theta_i$ like
\begin{align}
m(\boldsymbol{\theta}_i) &= \mu_{f_i \rightarrow \theta_i} \prod_{j \in \mathcal{D}_i} \mu_{f_{ji}  \rightarrow \theta_i} \nonumber \\
 &= \mu_{f_i \rightarrow \theta_i} \prod_{j \in \mathcal{D}_i} \int f_{ji} \  \mu_{\theta_j \rightarrow f_{ji}} d\boldsymbol{\theta}_j 
\label{eq:mp_1}
\end{align}
which are determined by marginalizing out all involved states except the state represented by the variable node of interest.
Since the factor graph has loops, a direct evaluation of \eqref{eq:mp_1} is not possible. Therefore we have to use an iterative message passing scheme where the belief of the state of agent $i$ at message passing iteration $u~\in~\{1,\dots,U\}$ is given by
\begin{align}
b^{(u)}(\boldsymbol{\theta}_i) \propto & \ b^{(0)}(\boldsymbol{\theta}_i) \prod_{a \in A} f(\bar{z}_{a,i}|\boldsymbol{\theta}_{i};\boldsymbol{\theta}_{a}) \nonumber \\
& \times \prod_{j \in \mathcal{D}_i} \int f(\underline{z}_{i,j}|\boldsymbol{\theta}_{i},\boldsymbol{\theta}_{j}) b^{(u-1)}(\boldsymbol{\theta}_j) d\boldsymbol{\theta}_j
\label{eq:BP_messagepassing}
\end{align}
using the beliefs of the agents ${j \in \mathcal{D}_i}$ from the previous iteration $b^{(u-1)}(\boldsymbol{\theta}_j)$. The initial belief $b^{(0)}(\boldsymbol{\theta}_i)$ is proportional to the prior PDF $f(\boldsymbol{\theta}_i)$. For the purpose of implementation, we can define the belief after incorporating the anchor measurements of the $i$th agent state as
\begin{equation}
\tilde{b}^{(0)}(\boldsymbol{\theta}_i) = b^{(0)}(\boldsymbol{\theta}_i) \prod_{a \in A} f(\bar{z}_{a,i}|\boldsymbol{\theta}_{i};\boldsymbol{\theta}_{a})
\end{equation}
since the states of the anchors have no dependency on other variables and are exactly known. With this definition \eqref{eq:BP_messagepassing} can be written as 
\begin{align}
b^{(u)}(\boldsymbol{\theta}_i) &\propto \tilde{b}^{(0)}(\boldsymbol{\theta}_i) \prod_{j \in \mathcal{D}_i} \int f(\underline{z}_{i,j}|\boldsymbol{\theta}_{i},\boldsymbol{\theta}_{j}) b^{(u-1)}(\boldsymbol{\theta}_j) d\boldsymbol{\theta}_j
\label{eq:BP_messag2}
\end{align} 
which leads to the same form as in \eqref{eq:mp_1}. We use a particle-based implementation \cite{meyer2015distributed_Loc} to approximate the calculations in \eqref{eq:BP_messag2} where the proposal densities for agent~$i$ and agent~$j$ at message passing iteration~$u$ are $\tilde{b}^{(0)}(\boldsymbol{\theta}_i)$ and $b^{(u-1)}(\boldsymbol{\theta}_j)$ respectively. Using $\tilde{b}^{(0)}(\boldsymbol{\theta}_i)$ in the particle-based implementation has the benefit that the particles are more condensed in the region of interest due to resampling after information from measurements to anchors is included. The estimation of the $i$th agent position and orientation is based on their marginal posterior PDF $f(\boldsymbol{\theta}_i|\mathbf{z})$ and is determined by the according minimum mean-square error (MMSE) estimator \cite{Kay1993}, i.e.,
\begin{equation}\label{eq:MMSE}
 \hat{\boldsymbol{\theta}}^{\text{MMSE}}_i = \int \boldsymbol{\theta}_i f(\boldsymbol{\theta}_i|\mathbf{z}) \text{d}\boldsymbol{\theta}_i
\end{equation}
where the agent marginal posterior PDF $f(\boldsymbol{\theta}_i|\mathbf{z})$ is approximated up to a normalization constant by the belief $b^{(U)}(\boldsymbol{\theta}_i)$.
This algorithm is an expansion of the sum-product algorithm for a wireless network (SPAWN) introduced in \cite{wymeersch2009cooperative} for static networks. In contrast to the traditional SPAWN, the orientation is included in the state representation. 

\subsection{Belief Propagation for discrete model}
\label{sec:BP_disc}
Based on the factor graph for the discrete inference model given in Fig.~\ref{fig:FG_disc}, we define the message passing scheme to approximate the marginal posterior PDFs. The difference to the case with continuous orientations is that we separate variable node $\theta_i$ into two variable nodes $x_i$ and $\Phi_i$. Those variable nodes have no direct dependency but influence each other via connecting factors. The orientation variable node $\Phi_i$ represents a discrete random variable which can take values of the finite set ${\mathcal{O}=\{\tilde{\varphi}_1, \tilde{\varphi}_2, \dots ,\tilde{\varphi}_{N_\mathcal{O}}\}}$ with the number of discrete orientations given as $N_\mathcal{O}$. The belief of the different states, which approximates the marginal posterior PDF, can be computed by multiplying all incoming messages at the different variable nodes respectively. For a better readability, we use the following short hand notation: The factor $f_{ij}~\triangleq~f(\underline{z}_{i,j}|\mathbf{x}_i,\Phi_{i},\mathbf{x}_j,\Phi_{j})$ represents the likelihood function with respect to the involved agent pair whereas factors $f_{i}~\triangleq~\sum_{\Phi_i \in \mathcal{O}} f(\mathbf{x}_i) p(\Phi_i) \prod_{a \in A} f(\bar{z}_{a,i}|\mathbf{x}_i,\Phi_{i};\mathbf{x}_a,\Phi_{a})$ and $p_{i}~\triangleq~\int f(\mathbf{x}_i) p(\Phi_i) \prod_{a \in A} f(\bar{z}_{a,i}|\mathbf{x}_i,\Phi_{i};\mathbf{x}_a,\Phi_{a}) dx_i$ represent prior information of the $i$th agent state and information from the anchor measurements. Additionally, we define $\mu_{x_i \rightarrow f_{ij}}$ as the message from variable node $\mathbf{x}_{i}$ to factor node $f_{ij}$, $\mu_{f_{ji}  \rightarrow x_i}$ as the message from factor node $f_{ji}$ to variable node $\mathbf{x}_{i}$, $\mu_{\Phi_i \rightarrow f_{ij}}$ as the message from variable node $\Phi_{i}$ to factor node $f_{ij}$ and $\mu_{f_{ji}  \rightarrow \Phi_i}$ as the message from factor node $f_{ji}$ to variable node $\Phi_{i}$. The belief of variable node $\mathbf{x}_{i}$ is given as 
\begin{align}
m(\mathbf{x}_i) = \ & \mu_{f_i  \rightarrow x_i} \prod_{j \in \mathcal{D}_i} \mu_{f_{ji}  \rightarrow x_i}  \nonumber \\
= \ & \mu_{f_i \rightarrow x_i} \prod_{j \in \mathcal{D}_i} \sum_{\Phi_i \in \mathcal{O}} \int \sum_{\Phi_j \in \mathcal{O}} f_{ji}  \nonumber \\
& \times   \mu_{x_j \rightarrow f_{ji}} \ \mu_{\Phi_j \rightarrow f_{ji}} \  d\mathbf{x}_j \label{eq:MP_2}
\end{align}
and the belief of variable node $\Phi_{i}$ as
\begin{align}
m(\Phi_i) = \ & \mu_{p_{i}  \rightarrow \Phi_i} \prod_{j \in \mathcal{D}_i} \mu_{f_{ji}  \rightarrow \Phi_i}  \nonumber \\
= \ & \mu_{p_{i}  \rightarrow \Phi_i}  \  \prod_{j \in \mathcal{D}_i} \iint \sum_{\Phi_j \in \mathcal{O}} f_{ji} \     \nonumber \\
&  \times \mu_{x_j \rightarrow f_{ji}} \  \mu_{\Phi_j \rightarrow f_{ji}}  d\mathbf{x}_j d\mathbf{x}_i.
\label{eq:MP_3}
\end{align}
by multiplying all incoming messages at the variable nodes respectively, which again are determined by marginalizing out all involved states except the state represented by the variable node of interest. Since the factor graph has loops, we can not calculate the messages at once but we have to approximate them iteratively. The approximate belief of the states $\mathbf{x}_i$ and $\Phi_i$ at message passing iteration $u$ can be written as
\begin{align}
b^{(u)}(\mathbf{x}_i) \rmv \propto \ & \sum_{\Phi_i \in \mathcal{O}} b^{(0)}(\mathbf{x}_i)b^{(0)}(\Phi_i) \prod_{a \in A} f(\bar{z}_{a,i}|\mathbf{x}_i,\Phi_{i};\mathbf{x}_a,\Phi_{a}) \nonumber \\
&\times \prod_{j \in \mathcal{D}_i} \int \sum_{\Phi_j \in \mathcal{O}} f(\underline{z}_{i,j}|\mathbf{x}_i,\Phi_{i},\mathbf{x}_j,\Phi_{j}) \nonumber \\
&\times b^{(u-1)}(\mathbf{x}_j)b^{(u-1)}(\Phi_j) d\mathbf{x}_j
\label{eq:BP_cont}
\end{align}
and
\begin{align}
b^{(u)}(\Phi_i) \rmv \propto \ & \int  \ b^{(0)}(\mathbf{x}_i)b^{(0)}(\Phi_i) \prod_{a \in A} f(\bar{z}_{a,i}|\mathbf{x}_i,\Phi_{i};\mathbf{x}_a,\Phi_{a}) \nonumber \\
& \times \prod_{j \in \mathcal{D}_i} \int \sum_{\Phi_j \in \mathcal{O}} f(\underline{z}_{i,j}|\mathbf{x}_i,\Phi_{i},\mathbf{x}_j,\Phi_{j}) \nonumber \\
& \times b^{(u-1)}(\mathbf{x}_j)b^{(u-1)}(\Phi_j) d\mathbf{x}_j d\mathbf{x}_i
\label{eq:BP_disc}
\end{align}
with $b^{(u)}(\Phi_i)$ being a PMF with $N_\mathcal{O}$ entries. The initial beliefs $b^{(0)}(\mathbf{x}_i)$ and $b^{(0)}(\boldsymbol{\Phi}_i)$ are proportional to the prior PDF $f(\mathbf{x}_i)$ and to the prior PMF $p(\boldsymbol{\Phi}_i)$, respectively. As in the previous section, we define a belief for $\mathbf{x}_i$ and $\Phi_i$ after incorporating the anchor measurements as
\begin{align}
\tilde{b}^{(0)}(\mathbf{x}_i) &= \sum_{\Phi_i \in \mathcal{O}} b^{(0)}(\mathbf{x}_i) b^{(0)}(\Phi_i) \prod_{a \in A} f(\bar{z}_{a,i}|\mathbf{x}_i,\Phi_{i};\mathbf{x}_a,\Phi_{a})  \\
\tilde{b}^{(0)}(\Phi_i) &= \int b^{(0)}(\mathbf{x}_i) b^{(0)}(\Phi_i) \prod_{a \in A} f(\bar{z}_{a,i}|\mathbf{x}_i,\Phi_{i};\mathbf{x}_a,\Phi_{a})  d\mathbf{x}_i
\end{align}
since the states of the anchors have no dependency on other variables and are exactly known. With this definition, \eqref{eq:BP_cont} and \eqref{eq:BP_disc} can be written as
\begin{align}
b^{(u)}(\mathbf{x}_i) \propto \ & \tilde{b}^{(0)}(\mathbf{x}_i) \sum_{\Phi_i \in \mathcal{O}} \prod_{j \in \mathcal{D}_i} \int \sum_{\Phi_j \in \mathcal{O}} f(\underline{z}_{i,j}|\mathbf{x}_i,\Phi_{i},\mathbf{x}_j,\Phi_{j}) \nonumber \\
 & \times b^{(u-1)}(\mathbf{x}_j) \ b^{(u-1)}(\Phi_j) d\mathbf{x}_j 
\label{eq:BP_cont2}
\end{align}
\begin{align}
b^{(u)}(\Phi_i)  \propto \ & \tilde{b}^{(0)}(\Phi_i) \int \prod_{j \in \mathcal{D}_i} \int \sum_{\Phi_j \in \mathcal{O}} f(\underline{z}_{i,j}|\mathbf{x}_i,\Phi_{i},\mathbf{x}_j,\Phi_{j}) \nonumber \\
& \times b^{(u-1)}(\mathbf{x}_j) \ b^{(u-1)}(\Phi_j) d\mathbf{x}_j  d\mathbf{x}_i
\label{eq:BP_disc2}\,.
\end{align}
The beliefs in \eqref{eq:BP_cont2} and \eqref{eq:BP_disc2} provide insights about the choice of proposal densities to efficiently represent regions of posterior PDF with high density.
It can be also observed how the belief of the orientation state of the previous iteration influences the belief of the position state and vice versa. Since we have a mixture of discrete and continuous probabilities, the marginalization to obtain the messages is performed by integrating over the continuous state $\mathbf{x}_j$ and the summation over the discrete state $\Phi_j$. The estimation of the $i$th agent position and orientation\footnote{Note that for orientation estimation, the mean of circular quantities is used.} is performed by their marginal PDF $f(\mathbf{x}_i|\mathbf{z})$ and marginal PMF $p(\Phi_i = \tilde{\varphi}_j|\mathbf{z})$, respectively, and determined by the according MMSE estimators, i.e.,
\begin{align}
\hat{\mathbf{x}}^{\text{MMSE}}_i &= \int \mathbf{x}_i f(\mathbf{x}_i|\mathbf{z}) \text{d}\mathbf{x}_i  \label{eq:MMSE_dis1} \\
\hat{\Phi}^{\text{MMSE}}_i &= \sum_{\tilde{\varphi}_j \in \mathcal{O}} \tilde{\varphi}_j p(\Phi_i = \tilde{\varphi}_j|\mathbf{z})\, .
\end{align}
where the agent marginal posterior PDF $f(\mathbf{x}_i|\mathbf{z})$ and PMF $p(\Phi_i \!\! = \!\! \tilde{\varphi}_j|\mathbf{z})$ are approximated up to normalization constants by the beliefs $b^{(U)}(\mathbf{x}_i)$ and $b^{(U)} (\Phi_i)$, respectively.

\captionsetup[subfigure]{labelformat=parens}
\begin{figure*}[t]
\centering
\subfloat[$\sigma = 1$ dB]
{\includegraphics[scale=1]{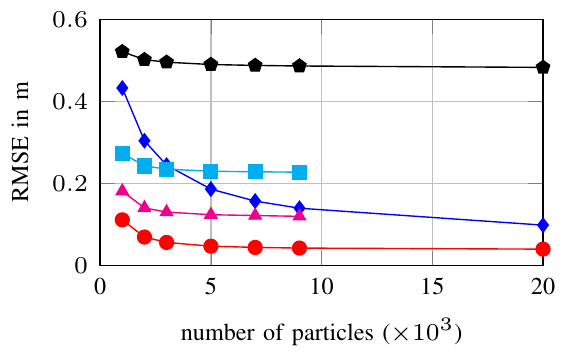}\label{fig:RMSE_1dB}}\hfill
\subfloat[$\sigma = 3$ dB]
{\includegraphics[scale=1]{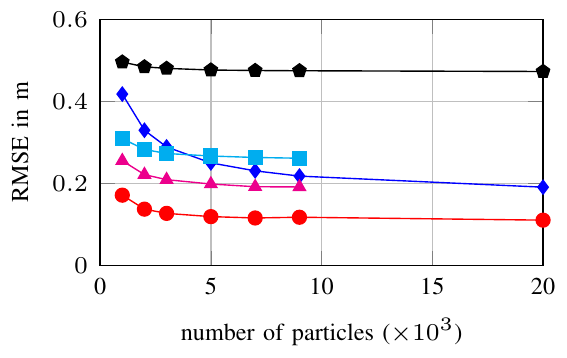}\label{fig:RMSE_3dB}}\hfill
\subfloat[$\sigma = 6$ dB]
{\includegraphics[scale=1]{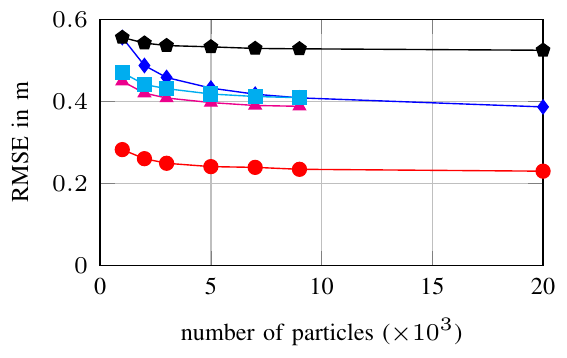}\label{fig:RMSE_6dB}}\hfill\\
\subfloat[$\sigma = 1$ dB]
{\includegraphics[scale=1]{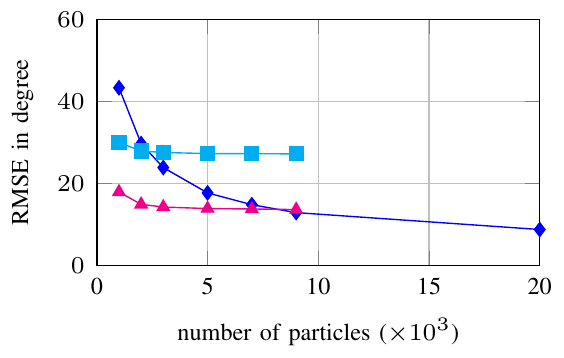}\label{fig:RMSE_angle_1dB}}\hfill
\subfloat[$\sigma = 3$ dB]
{\includegraphics[scale=1]{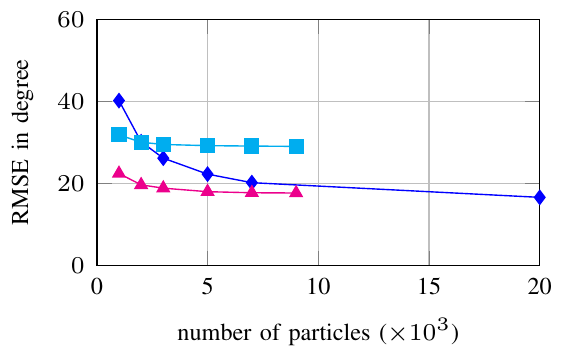}\label{fig:RMSE_angle_3dB}}\hfill
\subfloat[$\sigma = 6$ dB]
{\includegraphics[scale=1]{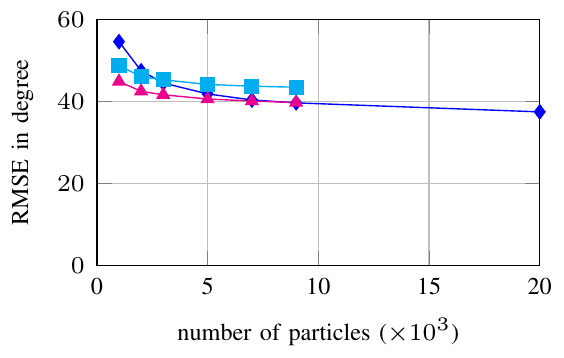}\label{fig:RMSE_angle_6dB}}\hfill\\
\includegraphics[scale=1]{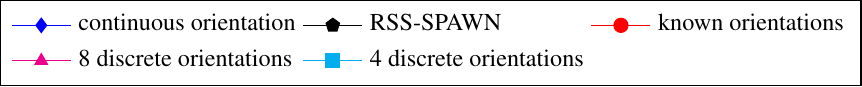}\hfill\\
\caption{Experiment 1: This figure shows the position RMSE and the RMSE of the orientation in dependency of the number of particles for 500 realizations of a scenario with 100 agents and 10 anchors for different uncertainties.}
\label{fig:2Drand100agents}
\end{figure*}

\section{Evaluation of Algorithms}
\label{sec:Results}
In this section, we study the performance of the two proposed methods based on simulations in a static 2D scenario. In addition, we compare it to the case where the orientation of the agents is known and to the classical SPAWN \cite{wymeersch2009cooperative,meyer2015distributed_Loc}, using an RSS measurement model without considering the orientation, similar to \cite{Jin2020_RSS_CL}. In the following we will refer to it as RSS-SPAWN. An example for why the orientation is neglected could be, if the antenna pattern is not known or if it is not known that the RSS depends on the orientation of the nodes.
The true agent and anchor positions are uniformly drawn for each realization on a support area of $5 \times 5\,$m. For the subsequent simulations, we use 100 agents and 10 anchors. In the following, we define the model parameters for model 1 ($k\!=\!1$). The path-loss parameters are given as $n_1 \! = \! 1$, $P_1 \! = \! -11$~dB with the reference distance $d_0$ chosen to be 0.1~m. The antenna pattern is given as a simple sinusoidal of the form of
\begin{equation}
\bar{\Psi}_1(\boldsymbol{\xi}_1,\phi) = \xi_{1,1} \text{cos}(\phi + \xi_{1,2})
\label{eq:antennaPattern}
\end{equation}
with the amplitude $\xi_{1,1}\!=\!3.36$~dB and the antenna orientation $\xi_{1,2}\!=\!0.11$~rad which shows the deviation of the maximum of the antenna pattern from the true orientation. In this section, anchors have a uniform antenna pattern with 0~dB gain. The resulting variation of the measured signal strength, which depends only on the orientation of two agents is given as
\begin{equation}
\Psi_1(\boldsymbol{\xi}_1,\phi_{i,j},\phi_{j,i}) = \xi_{1,1}\text{cos}(\phi_{i,j} + \xi_{1,2}) + \xi_{1,1}\text{cos}(\phi_{j,i} + \xi_{1,2})
\end{equation}
To determine the marginal posterior PDFs of the agent states, we use a particle-based belief propagation implementation \cite{meyer2015distributed_Loc} to cope with multimodal distributions. 
%
\subsection{Experiment 1: Random Orientation}
\label{sec:exp1}
In this analysis, we show for the four previously mentioned methods, how the accuracy of the positioning performance depends on the number of particles. This corresponds to the achieved resolution of the continuous state. The orientations of the agents are drawn uniformly in the range of $[0,2\pi)$. For the method that uses the discrete inference model, we perform the evaluation with two different sets of orientations which consist of four and eight equally-spaced members in the range of $[0,2\pi)$, respectively. For a set of four, it results in ${\mathcal{O} = \{0,\pi/2,\pi,-\pi/2 \}}$. The results are shown in Fig.~\ref{fig:2Drand100agents} for three different measurement uncertainties ${\sigma = [1,3,6]}$~dB and 500 simulation runs per point. We can see that using the discrete inference model has the highest benefit at a low number of particles. The reason for this behavior is that, for a low number of particles, the resolution of the joint state for continuous orientations is too low to correctly estimate the position and the orientation of the agents. For discrete orientations, this effect is not so severe since the resolution of the orientation is determined by the discrete orientations. Therefore, if the continuous state of the agent positions is already well enough represented with particles, increasing the number of particles will only lead to a minor decrease of the positioning error. For that reason, we neglect the evaluation of 20000 particles for the discrete inference model since the results in Fig.~\ref{fig:2Drand100agents} show already a converged behaviour of the RMSE at a lower number of particles. Above some number of particles, the method using the continuous inference model outperforms the discrete one. This is due to the fact that the resolution of the orientation of the discrete method is limited by the discrete orientations if the true orientations are not in the set of discrete orientations ${\varphi_i \notin \mathcal{O}}$. Therefore using eight discrete orientations has a much better performance in terms of positioning and orientation accuracy compared to using four discrete orientations since eight equally spaced orientations approximate the continuous orientation space better. By using different measurement uncertainties, we can better see the difference in performance of the investigated methods. At this point we want to mention that reducing the measurement uncertainty but still neglecting the orientation, has nearly no impact on the performance in terms of RMSE. This is due to the bias that occurs when generating the data with an antenna pattern
\subsection{Complexity Analysis}
\label{sec:CompAnalysis}
\begin{figure}[t]
\centering
\includegraphics[scale=1]{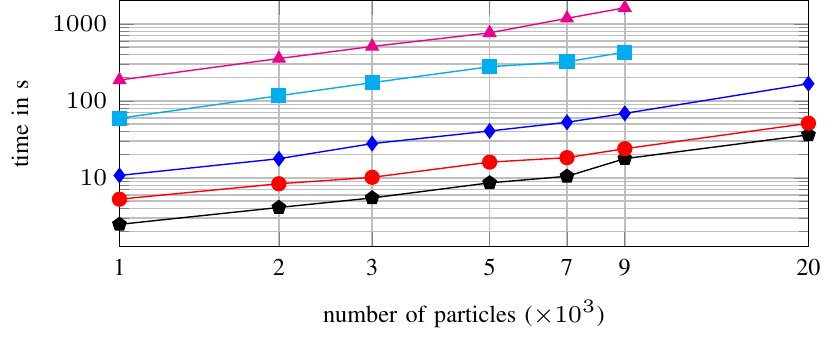}\hfill
\includegraphics[scale=1]{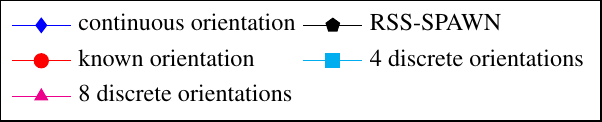}\hfill\\
\caption{Complexity analysis: This figure shows the calculation time for Experiment 1 averaged over all 1500 simulation runs, which consists of all three tested measurement uncertainties. It can be seen how the complexity in terms of the calculation time depends on different methods and on the number of particles.}
	\label{fig:measTime}
\end{figure}

In this section we give an estimate of the complexity of the investigated methods depending on the number of agents $|\mathcal{C}|$ and the number of particles $N_P$. For the continuous inference model, the complexity is of the order ${O_c(|\mathcal{C}|^2 \times N_P)}$ whereas for the discrete inference model, we have an additional quadratic dependency on the number of discrete orientations $|\mathcal{O}|$ which results in ${O_d(|\mathcal{C}|^2 \times N_P \times |\mathcal{O}|^2)}$. We validate this estimate by measuring the computation time of Experiment~1. The results are shown in Fig.~\ref{fig:measTime}. We can see the strong increase in computation time for the discrete methods compared to the continuous ones. The ratio of the computation time between different methods is approximately the same as calculated with our assessment. The discrepancy regarding known orientation compared to the RSS-SPAWN implementation is due to some extra calculations which include transformations from Cartesian to polar coordinates. The difference in-between continuous and known orientation can be explained with the additional orientation estimation, which leads to an increased amount of calculations.
%
%
\subsection{Experiment 2: Increased Number of Orientations}
\label{sec:complexity}
\begin{figure}[t]
\centering
\subfloat[Position RMSE]
{\includegraphics[scale=1]{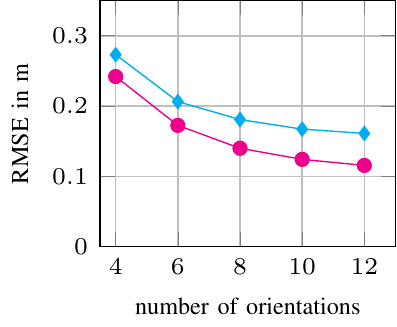}\label{fig:varO_posRMSE_1dB}}
\subfloat[Orientation RMSE]
{\includegraphics[scale=1]{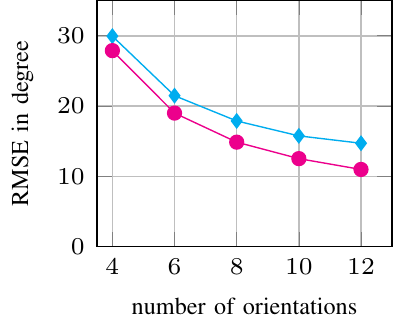}\label{fig:varO_angleRMSE_1dB}}\hfill
\includegraphics[scale=1]{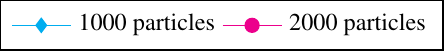}\hfill\\
\caption{Experiment 2: This figure shows the position RMSE and the RMSE of the orientation in dependency of the number of discrete orientations for 500 realizations of a scenario with 100 agents and 10 anchors. The standard deviation of the log-normal shadowing is $\sigma$~=~1~dB.}
\label{fig:RMSE_varO}
\end{figure}

We show in Section \ref{sec:exp1} that for the discrete inference model, the number of particles has only a small impact on the localization performance compared to the fully continuous algorithm. Increasing the number of possible orientations on the other hand, yields an improvement of the accuracy of estimating position and orientation since more hypotheses can be evaluated. The result for that analysis is depicted in Fig.~\ref{fig:RMSE_varO}. It shows the dependency of the positioning and orientation RMSE on the number of orientations for 1000 and 2000 particles using $\sigma$~=~1~dB and evaluated for 500 simulation runs.
The discrete orientations are equally spaced in the range of $[0,2\pi)$. A disadvantage of increasing the number of orientations is that it results in a higher computational complexity as shown in Section \ref{sec:CompAnalysis}.
%
%
\subsection{Experiment 3: Prior Knowledge of Orientation}
\begin{figure*}[t]
\centering
\subfloat[$\sigma = 1$ dB]
{\includegraphics[scale=1]{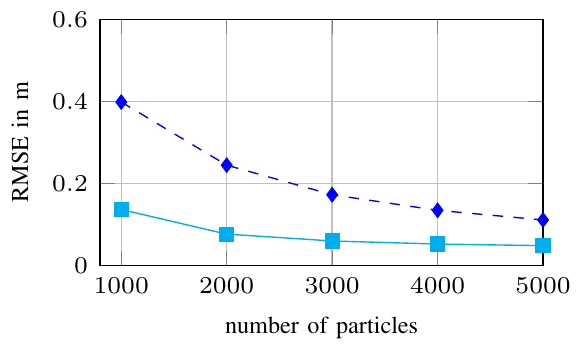}\label{fig:RMSE_1dB_O4}}\hfill
\subfloat[$\sigma = 3$ dB]
{\includegraphics[scale=1]{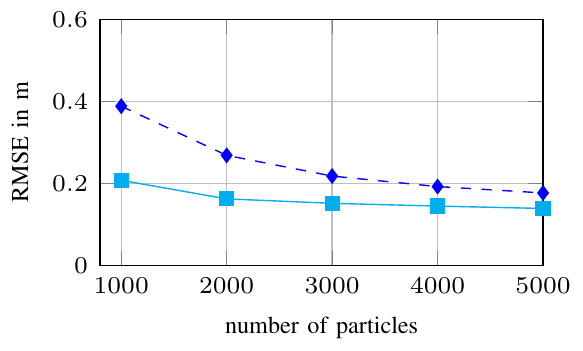}\label{fig:RMSE_3dB_O4}}\hfill
\subfloat[$\sigma = 6$ dB]
{\includegraphics[scale=1]{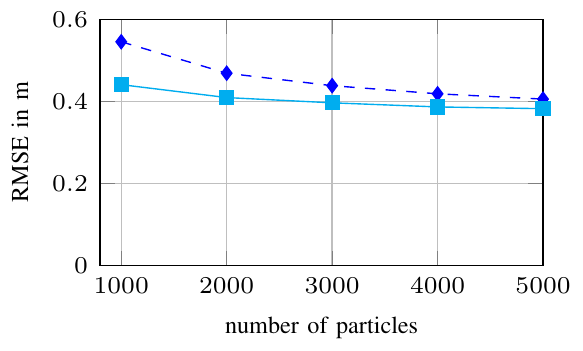}\label{fig:RMSE_6dB_O4}}\hfill \\
\subfloat[$\sigma = 1$ dB]
{\includegraphics[scale=1]{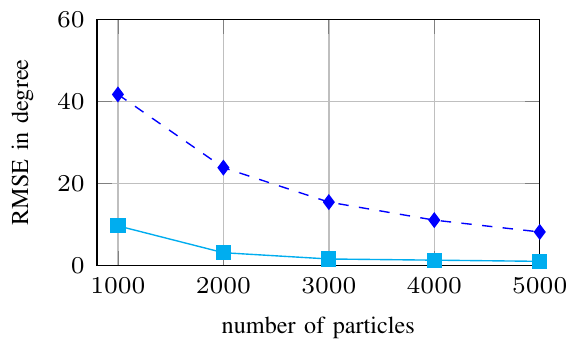}\label{fig:RMSE_angle_1dB_O4}} \hfill
\subfloat[$\sigma = 3$ dB]
{\includegraphics[scale=1]{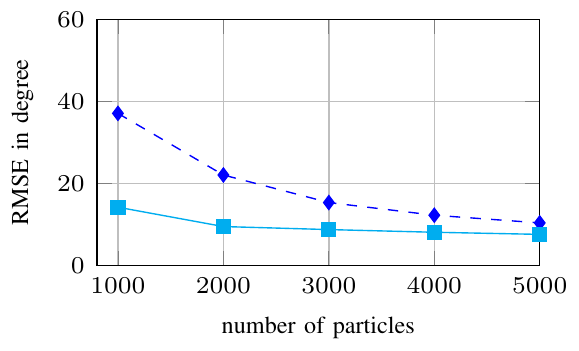}\label{fig:RMSE_angle_3dB_O4}}\hfill
\subfloat[$\sigma = 6$ dB]
{\includegraphics[scale=1]{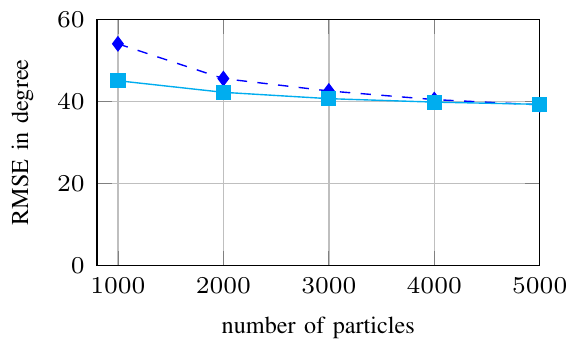}\label{fig:RMSE_angle_6dB_O4}}\hfill \\
\includegraphics[scale=1]{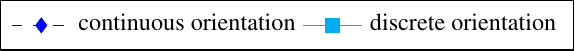}\hfill\\
\caption{Experiment 3: This figure shows the position RMSE and the RMSE of the orientation in dependency of the number of particles for 500 realizations of a scenario with 100 agents and 10 anchors for different uncertainties. The orientations of the agents are chosen such that they are in the set of four possible discrete hypothesis which are used in the discrete orientation estimation indicated in cyan. The four possible hypothesis are also used as a prior for the continuous orientation estimation which is indicated in blue.}
    \label{fig:2Drand100agents_trueO}
\end{figure*}

For this investigation, we look at the case where prior information about the orientation of the agents is available. This could be the case if, due to the geometry of the scenario, only certain orientations of the agents are possible. An example could be mounting the agents on parallel shelfs or on other objects. For the simulations, we draw the orientations of the 100 agents at random out of the set of four possible orientations $\mathcal{O} = \{0,\pi/2,\pi,-\pi/2 \}$ where $0$~rad corresponds to the positive x-axis and $\pi/2$~rad to the positive y-axis. Those orientations are used as prior knowledge for both described algorithms. This means that the proposal density for the continuous orientations is a discrete uniform distribution, which consists of $\mathcal{O}$. Using the discrete inference model results in only testing true possible combinations of orientations. In Fig. \ref{fig:2Drand100agents_trueO}, we show the RMSE given position and orientation estimates for a varying number of particles and different uncertainties of the measurement model. We can see that the algorithm which uses the discrete inference model outperforms the algorithm using the continuous inference model especially for a small measurement uncertainty and a small number of particles.
%
\captionsetup[subfigure]{labelformat=parens}
\section{Evaluation with Measured Data}
\label{sec:Meas_Results}
In this section, we evaluate the proposed algorithms with data recorded at a measurement campaign in a library at TU Graz. For this purpose, we extend the algorithms to work for 3D scenarios. The used nodes are electronic shelf labels. To validate the results for the measurement campaign, we compare them to a synthetic scenario with the same geometrical structure and model parameters as for the real scenario. The definition of the scenario, which is used in both cases, is given in the Section~\ref{sec:scenario}. The model selection for the antenna pattern as well as the parameter estimation for path-loss and antenna pattern is given in Section~\ref{sec:detAP}. The results for the synthetically generated data are shown in Section~\ref{subsec:SYNTH} whereas the results for the measurement campaign are shown in Section~\ref{subsec:MEAS}.
\begin{figure}[t]
\centering
\includegraphics[scale=1]{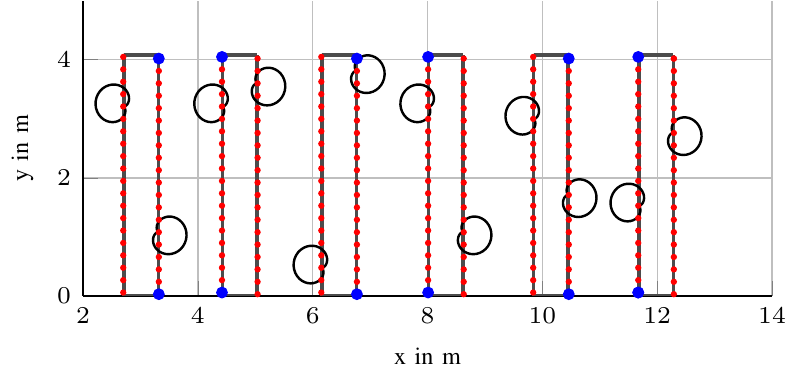}\hfill
\includegraphics[scale=1]{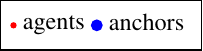}\hfill\\
\caption{This figure shows the top view of the scenario. Each marker corresponds to four different heights, equally spaced between 0.8~m and 1.8~m. All nodes on one side of the shelf have the same orientation which is indicated by the antenna pattern in black. Anchors are placed at the lowest and highest hight of the scenario at the same x and y coordinates as indicated in the figure.}
	\label{fig:synthScen}
\end{figure}
\subsection{Scenario}
\label{sec:scenario}
The scenario is a section of a library where the nodes are mounted on six shelfs as shown in Fig. \ref{fig:synthScen}. We use 960 nodes which are equally spaced on four different heights, ranging from 0.8~m to  1.8~m. The spacing in the y-direction is 0.2~m, resulting in 20 nodes per side of the shelf and per height. The width of the shelfs is 0.6~m and the width of the corridors 1.1~m. The scenario consists of 936 agents and 24 anchors. All nodes on one side of the shelf have the same orientation, which is visualized in Fig.~\ref{fig:synthScen} with antenna patterns in black. The set of true orientations is $\mathcal{O}_{\text{true}} = \{0,\pi\}$. The support area, which is used as a uniform prior distribution for each agent position state, is given as $x_{\text{prior}} = [0.7, 14.3]$~m, $y_{\text{prior}} = [-2, 6]$~m, $z_{\text{prior}} = [0.8, 1.8]$~m.
%
\subsection{Determination of Antenna Pattern}
\label{sec:detAP}
In this section, we explain how the measurement model is selected and how the model parameters are estimated. We assume that the positions and orientations of the nodes are known. We use a parametric representation of the antenna pattern with an unknown number of parameters. The measurements conditioned on the model $M\!\!=\!\!k$ can be seen in \eqref{eq:measModelAP}.
The set of used nodes for the parameter estimation is indicated as $\mathcal{S}$, where $|\mathcal{S}|$ is the number of used nodes. The measurements in-between nodes are indicated as $z_{i,j}$ with $i \in \mathcal{S}$ and $j \in \mathcal{K}_i$ with $\mathcal{K}_i \subseteq \mathcal{S}$ being the set of nodes that cooperate with node $i$. 
Since the positions and orientations of the involved nodes are known, the likelihood function in \eqref{eq:lhf} does not depend on $\boldsymbol{\theta}$ and can be written in the form of $f(z_{i,j}|\boldsymbol{\vartheta}_k,M\!\!=\!\!k)$. 
For a known model, the parameters are estimated by maximizing the likelihood function as
\begin{equation}
\boldsymbol{\hat{\vartheta}}_k = \argmax_{\boldsymbol{\vartheta}_k} \prod_{i = 1}^{|\mathcal{S}|} \prod_{j \in \mathcal{K}_i} f(z_{i,j}|\boldsymbol{\vartheta}_k,M\!\!=\!\!k)
\label{eq:argmax}
\end{equation} 
We solve \eqref{eq:argmax} in an iterative manner where one iteration consists of first estimating the path-loss parameters, then the parameters for the antenna pattern, and at last the shadowing standard deviation. This is repeated until some convergence criterion is reached.
To determine which model $M$ out of a set of possible models is more likely to have generated the data $\mathbf{z}$, we are interested in the posterior of the models given as 
\begin{equation}
f(M|\mathbf{z}) = \frac{f(\mathbf{z}|M) f(M)}{f(\mathbf{z})}.
\label{eq:postModel}
\end{equation}
The evidence of all models $f(\mathbf{z})$ is hard to calculate. Therefore we are only looking at the ratio between the posterior distributions of the models which has the benefit that the  evidence of all models cancels out. This is the so called odds ratio \cite{buckland1997model,von2014bayesian}.
The odds ratio $O_{k,j}$ between two models $M\!\!=\!\!k$ and $M\!\!=\!\!j$ with $k,j \in \mathcal{M}$ is defined as 
\begin{equation}
O_{k,j} = \frac{f(M\!\!=\!\!k|\mathbf{z})}{f(M\!\!=\!\!j|\mathbf{z})} = \frac{f(M\!\!=\!\!k)}{f(M\!\!=\!\!j)} \frac{f(\mathbf{z}|M\!\!=\!\!k)}{f(\mathbf{z}|M\!\!=\!\!j)}
\label{eq:odds}
\end{equation}
and favours model $k$ if $O_{k,j} > 1$. Using marginalization, we can rewrite the likelihood given model $M\!\!=\!\!k$ in \eqref{eq:odds} as
\begin{equation}
f(\mathbf{z}|M\!\!=\!\!k) = \int f(\mathbf{z}|\boldsymbol{\vartheta}_k,M\!\!=\!\!k) f(\boldsymbol{\vartheta}_k) d\boldsymbol{\vartheta}_k
\end{equation}
with $f(\boldsymbol{\vartheta}_k)$ being the prior PDF of the parameters of model $k$. Assuming that the prior is flat in the region of interest, we can approximate that integral using the Bayesian information criterion (BIC) \cite{neath2012bayesian}. This leads to
\begin{equation}
\text{ln} f(\mathbf{z}|M\!\!=\!\!k) \approx \text{ln} f(\mathbf{z}|\hat{\boldsymbol{\vartheta}}_k,M\!\!=\!\!k) - \frac{{N}_{\vartheta_k}}{2} \text{ln} N_z
\label{eq:BIC}
\end{equation}
with the number of parameters $N_{\vartheta_k}$, the number of measurements $N_z$ and the parameters that maximize the likelihood function as $\hat{\boldsymbol{\vartheta}}_k$, which can be calculated by solving \eqref{eq:argmax}.
\begin{figure}[t]
\centering
\subfloat[Measured data]
{\includegraphics[scale=1]{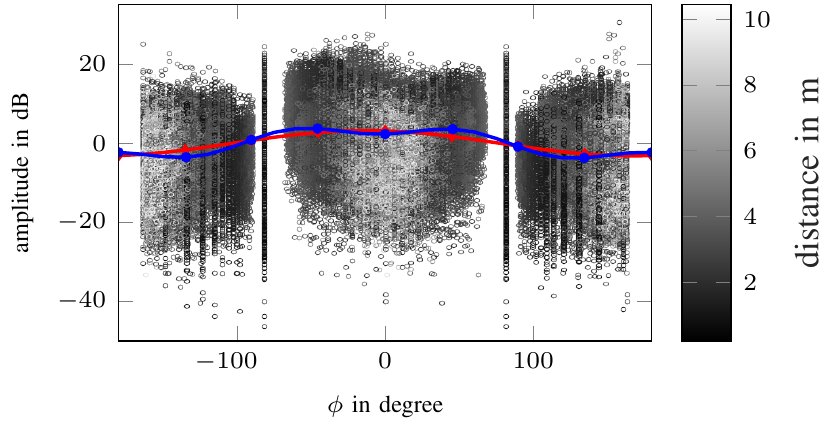}\label{fig:aziMeas}}\hfill
\subfloat[Synthetic data generated with $M\!\!=\!\!2$]
{\includegraphics[scale=1]{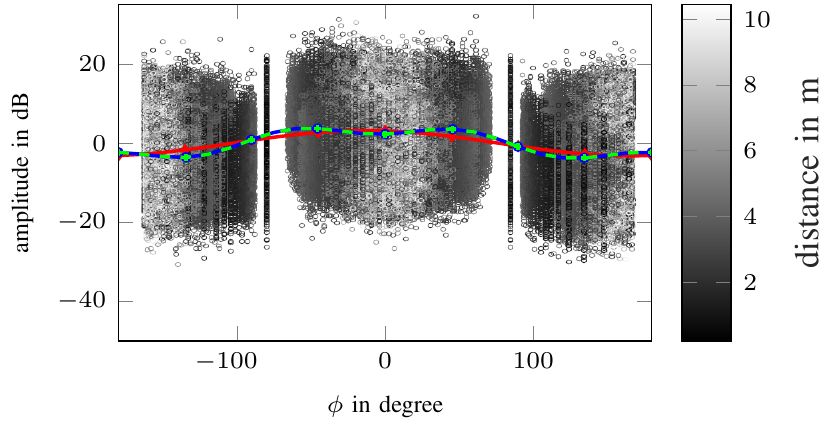}\label{fig:aziSynthComplex}}\hfill
\includegraphics[scale=1]{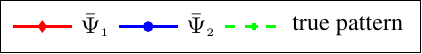}\hfill\\
\caption{This figure shows measured and synthetic data with subtracted path-loss in dependency of the angle in-between the nodes. The distance between the nodes is indicated in gray.}
\end{figure}
With the BIC given in \eqref{eq:BIC}, we can calculate the odds ratio in \eqref{eq:odds}, assuming that all models are equally probable. Note that since the models differ only in terms of the antenna pattern, finding the most probable model is equivalent to finding the most probable antenna pattern (see \eqref{eq:measModelAP}).
We will use this criterion to compare two different antenna models $\bar{\Psi}_k$ with one another. The models are given as
\begin{align}
\bar{\Psi}_1(\boldsymbol{\xi}_1,\phi) = & \ \xi_{1,1} \text{cos}(\phi + \xi_{1,2}) \nonumber \\
\bar{\Psi}_2(\boldsymbol{\xi}_2,\phi) = & \ \xi_{2,1} \text{cos}(\phi + \xi_{2,2})  + \xi_{2,3}\text{cos}(3 \phi + \xi_{2,4})
\label{eq:models}
\end{align}
where the overall influence on the measurement can be calculated using \eqref{eq:2wayAP}. The models are chosen by looking at the data in Fig.~\ref{fig:aziMeas}. Note that $\bar{\Psi}_1$ is fully included in $\bar{\Psi}_2$ and that $\bar{\Psi}_2$ is a harmonic extension of $\bar{\Psi}_1$. Evaluating the BIC shows with overwhelming evidence that model 2 is favoured, i.e. $\bar{\Psi}_2$ models the data more precisely. The parameters for the antenna pattern models given in \eqref{eq:models} are determined as  {$\hat{\boldsymbol{\xi}}_1=[3.22, 0.22]^\text{T}$} and {$\hat{\boldsymbol{\xi}}_2=[3.76, 0.13, -1.47, 0.28]^\text{T}$}.
We further investigate if it is possible to extract the correct antenna pattern from synthetic data. For that purpose, we generate synthetic measurements according to $M\!\!=\!\!2$ and $\hat{\boldsymbol{\vartheta}}_2$, with ${\hat{n}_2=1.09}$, ${\hat{P}_2=-9.18}$~dB for ${d_0=0.1}$~m and $\hat{\sigma}_2=5.77$~dB, and estimate the model parameters based on the generated measurements. The synthetic measurements as well as the generating model (green) and the estimated model (blue) can be seen in Fig.~\ref{fig:aziSynthComplex}. Evaluating the BIC for $M\!\!=\!\!1$ and $M\!\!=\!\!2$ given this data set, results again in a strong favouring of model 2 which is expected since the data is generated with model 2. The estimated parameters for model 2 are very close to the generating model parameters. If less data are available and if the data have a sufficiently large variance, model 1 would be preferred over model 2. We want to mention at this point that the estimated model for the antenna pattern based on the measured data is not the actual antenna pattern of the nodes but an effective pattern that depends also on the environment in which the measurements are performed. We estimate the model parameters based on all available measurements (global parameters) but it is also possible to determine them based on a subset of nodes (see Section~\ref{subsec:nodeSubset}). We do not look into further detail regarding at which number and placement of nodes the antenna pattern can be estimated correctly. 
Since model 2 ($M\!\!=\!\!2$) is favored, we use model 2 and $\hat{\boldsymbol{\vartheta}}_2$ for the following analysis. 
%
\subsection{Synthetic Results}
\label{subsec:SYNTH}
To gain insight how the algorithms perform in such a structured scenario, we evaluate 50 simulation runs regarding measurements between nodes. The synthetic data is generated with model 2 and $\hat{\boldsymbol{\vartheta}}_2$. The model parameters for the evaluations are estimated with regard to \eqref{eq:argmax} based on the synthetically generated data. For the discrete inference model, we test four orientations $\mathcal{O} = \{0,\pi/2,\pi,-\pi/2 \}$, which correspond to prior knowledge that the nodes can only be mounted on shelfs such that the orientation is normal to the possible segments of the shelfs (see Fig.~\ref{fig:synthScen}). We use 1000 particles to represent the state of each agent. For the continuous inference model, we use different proposal densities for the orientation. The first one is a uniform distribution in the range of $[-\pi, \pi)$ and the second one a discrete uniform distribution which consists of $\mathcal{O}$. For each proposal density, we perform 50 simulation runs with a number of particles $N_P$ of 1000 and 4000. To achieve a proper evaluation, we compare the proposed methods to the case where the orientation of the agents is exactly known and to the case where the orientation is completely neglected which corresponds to the classical RSS-SPAWN implementation. The latter one coincides to a model mismatch which would be the case if the antenna pattern is not known or if it is not known that the RSS depends on the orientation of the nodes. For both comparisons, we use 1000 and 4000 particles. 

\begin{figure}[t]
\subfloat[CF of position error]
{\includegraphics[scale=1]{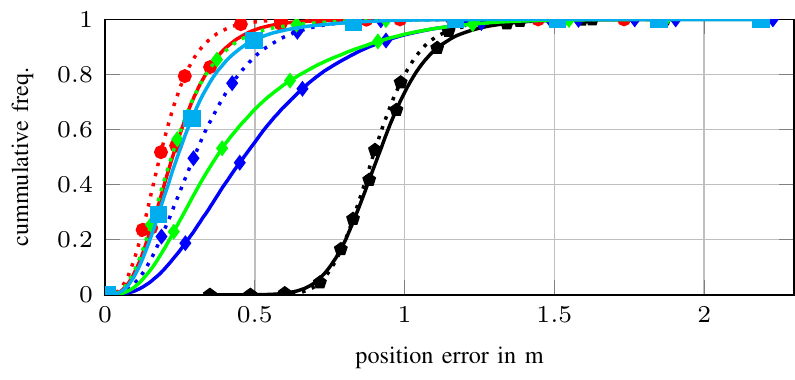}\label{fig:synthScenPosCDF}}\hfil
\subfloat[CF of absolute orientation error]
{\includegraphics[scale=1]{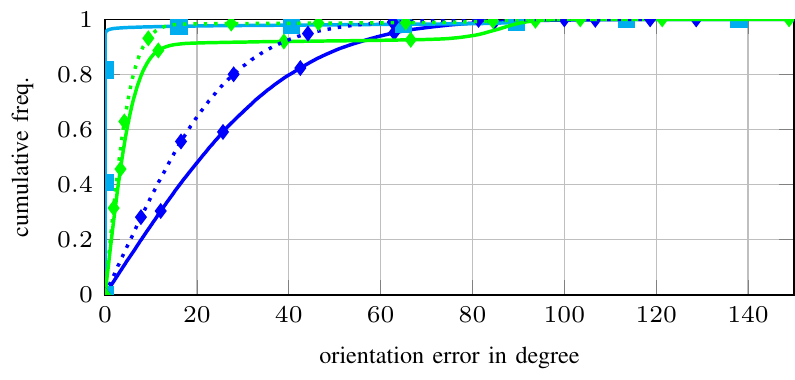}\label{fig:synthScenAngleCDF}}\hfil
\centering
\includegraphics[scale=1]{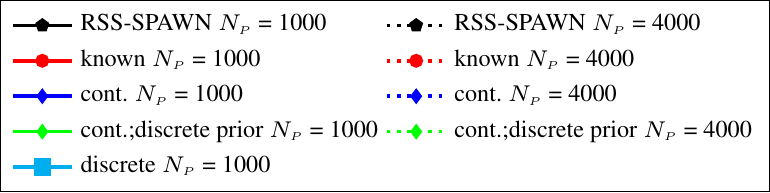}\hfill\\
\caption{Synthetic: This figure shows the CF of the position error and the absolute orientation error of the 3D synthetic scenario for different evaluations and different numbers of particles. The abbreviation 'cont.' stands for the continuous orientation state where the proposal density for the orientation is a uniform distribution in the range of $[-\pi, \pi)$. In the case of 'cont.; discrete prior', the proposal density for the orientations is a discrete uniform distribution that consists of $\mathcal{O}$.}
    \label{fig:synthScenCDF}
    \vspace*{-2mm}
\end{figure}

In Fig.~\ref{fig:synthScenCDF}, we show the cumulative frequency (CF) of the position error in meters and the CF of the absolute orientation error in degree. It can be seen that estimating position and orientation leads to a significant improvement of the positioning accuracy compared to neglecting the orientation. An interesting aspect of this result is that prior information about the orientation has not so much impact on the positioning performance of the algorithm that uses the continuous inference model. The main influence is, as mentioned in Section~\ref{sec:exp1}, on the number of particles which corresponds to the resolution of the state. This behaviour cannot be seen for the angular error. Here, using prior knowledge of the orientation leads to an advantage regardless of the number of particles. For the same resolution of the position state, the algorithm that uses the discrete inference model clearly outperforms the other approaches. It is also very close the ground truth with fixed orientation. Compared to the method with the discrete inference model, the algorithm with the continuous inference model, 4000 particles and prior information about the orientation has a slightly better performance in terms of the positioning error. Regarding the orientation error, using the discrete inference model results in much more correctly estimated orientations but also in more outliers which reduces the overall performance. A summary of the results is given in Table~\ref{tb:results_synth}. In addition to the evaluation with known orientation, we perform another evaluation of the algorithm with the true model parameters and known orientation, which is called 'gold standard'. The results show that there is nearly no difference in terms of the positioning RMSE in between the two analyses. It indicates that the model parameter estimation works properly. Note that if the orientation is neglected, the second part of \eqref{eq:lhf_param} is omitted. When estimating the parameters using \eqref{eq:argmax}, it results in a larger shadowing variance since it has to account for the variations caused by the antenna pattern.

\begin{table}[t]
  \caption{Synthetic: RMSE for position and orientation for synthetic results based on 50 simulation runs. The abbreviation ``cont.'' stands for continuous and ``prior cont'' for using prior information about the orientations in the continuous representation.}
\begin{center}
  \begin{tabular}{ l | l| r | r | r | r |}
Orientation & $N_P$ & \multicolumn{2}{|c|}{24 anchors} & \multicolumn{2}{|c|}{48 anchors} \\
\hline
     &  & $\text{RMSE}_{\text{p}}$ &   $\text{RMSE}_{\text{o}}$ &  $\text{RMSE}_{\text{p}}$ & $\text{RMSE}_{\text{o}}$    \\ \hline
RSS-SPAWN	  & 1000 & 0.94 m &    -       & 0.91 m &    -       \\ \hline
RSS-SPAWN	  & 4000 & 0.92 m &    -       & 0.89 m &    -       \\ \hline
gold standard & 1000 & 0.28 m &    -       & 0.25 m &    -       \\ \hline
gold standard & 4000 & 0.22 m &    -       & 0.20 m &    -       \\ \hline
known	      & 1000 & 0.28 m &    -       & 0.25 m &    -       \\ \hline
known	      & 4000 & 0.22 m &    -       & 0.20 m &    -       \\ \hline
discrete	  & 1000 & 0.31 m & 12$^\circ$ & 0.28 m & 10$^\circ$ \\ \hline
cont.         & 1000 & 0.58 m & 32$^\circ$ & 0.53 m & 31$^\circ$ \\ \hline
cont.         & 4000 & 0.36 m & 23$^\circ$ & 0.33 m & 22$^\circ$ \\ \hline
prior cont.   & 1000 & 0.53 m & 24$^\circ$ & 0.47 m & 23$^\circ$ \\ \hline
prior cont.   & 4000 & 0.28 m & 11$^\circ$ & 0.25 m &  9$^\circ$ \\ \hline
 \end{tabular}
\end{center}
\label{tb:results_synth}
\end{table}

\begin{table}[t]
  \caption{Measurement: RMSE for position and orientation for measured results. The abbreviation ``cont.'' stands for continuous and ``prior cont.'' for using prior information about the orientations in the continuous representation.}
\begin{center}
  \begin{tabular}{ l | l | r | r | r | r |}
Orientation & $N_P$ & \multicolumn{2}{|c|}{24 anchors} & \multicolumn{2}{|c|}{48 anchors} \\
\hline
   &  &  $\text{RMSE}_{\text{p}}$ &   $\text{RMSE}_{\text{o}}$ &  $\text{RMSE}_{\text{p}}$ & $\text{RMSE}_{\text{o}}$    \\ \hline
RSS-SPAWN	 & 1000 & 1.16 m &    -       & 1.12 m &    -       \\ \hline
RSS-SPAWN	 & 4000 & 1.14 m &    -       & 1.12 m &    -       \\ \hline
known	     & 1000 & 0.75 m &    -       & 0.70 m &    -       \\ \hline
known	     & 4000 & 0.71 m &    -       & 0.70 m &    -       \\ \hline
discrete     & 1000 & 0.85 m & 51$^\circ$ & 0.82 m & 53$^\circ$ \\ \hline
cont.        & 1000 & 0.88 m & 55$^\circ$ & 0.90 m & 58$^\circ$ \\ \hline
cont.        & 4000 & 0.77 m & 48$^\circ$ & 0.78 m & 48$^\circ$ \\ \hline
prior cont.  & 1000 & 0.93 m & 60$^\circ$ & 0.89 m & 57$^\circ$ \\ \hline
prior cont.  & 4000 & 0.83 m & 49$^\circ$ & 0.80 m & 49$^\circ$ \\ \hline
 \end{tabular}
\end{center}
\label{tb:results_meas}
\end{table}

\subsection{Measurement Results}
\label{subsec:MEAS}
\begin{figure}[t]
\subfloat[CF of position error]
{\includegraphics[scale=1]{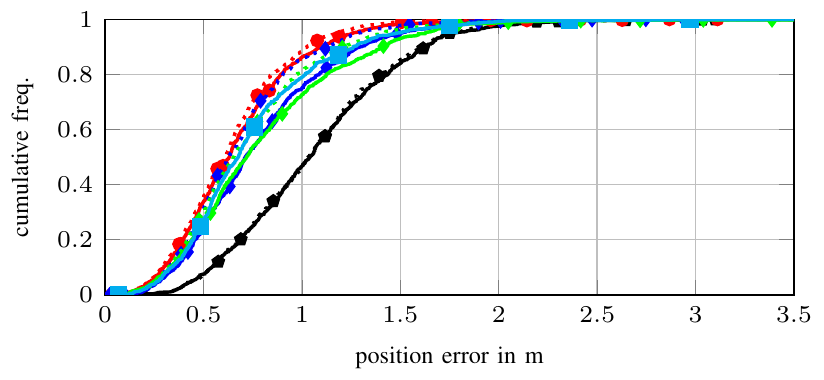}\label{fig:measScenPosCDF}}
\hspace{1cm}
\subfloat[CF of absolute orientation error]
{\includegraphics[scale=1]{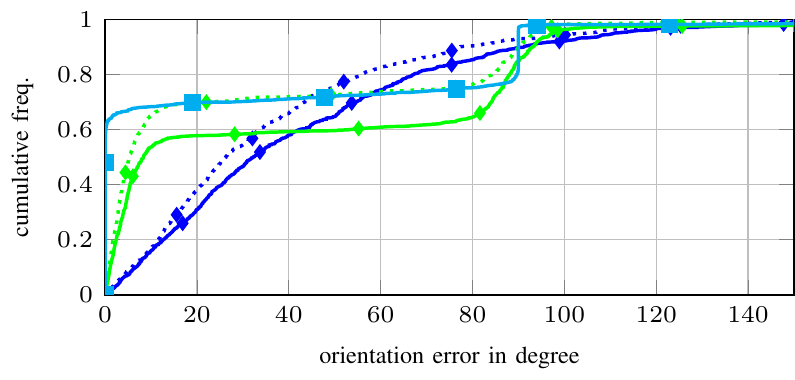}\label{fig:measScenAngleCDF}}\hfill
\centering
\includegraphics[scale=1]{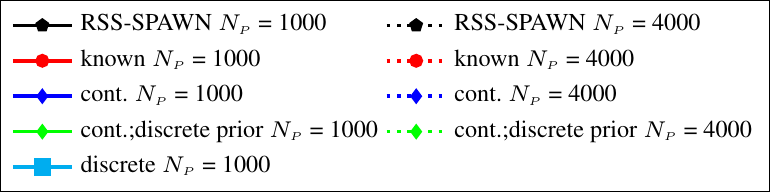}\hfill\\
\caption{Measurement: This figure shows the CF of the position error and the absolute orientation error for different evaluations and different numbers of particles. The abbreviation 'cont.' stands for the continuous orientation state where the proposal density for the orientation is a uniform distribution in the range of $[-\pi, \pi)$. In the case of 'cont.; discrete prior', the proposal density for the orientations is a discrete uniform distribution that consists of $\mathcal{O}$.}
    \label{fig:measScenCDF}
\end{figure}

In this section, we evaluate the proposed algorithms with real data resulting from the measurement campaign introduced in Section \ref{sec:scenario}.
The path-loss model and the antenna pattern are estimated as explained in Section \ref{sec:detAP}. We perform the same evaluation of the algorithms as for the synthetic scenario. For the discrete inference model, we use 1000 particles and four orientations ${\mathcal{O} = \{0,\pi/2,\pi,-\pi/2 \}}$. Regarding the method which utilizes the continuous inference model, we use a uniform distribution in the range of $[-\pi,\pi)$ and a discrete uniform distribution that consists of $\mathcal{O}$ as proposal densities. Each proposal density is evaluated with 1000 and 4000 particles. Additionally, we compare to the case of known orientation and the RSS-SPAWN which completely neglects the orientation. The agent state is represented with 1000 and 4000 particles. The CF for the positioning error as well as for the absolute orientation error are given in Fig. \ref{fig:measScenCDF}. Note that the largest outlier is at 5.5~m. The results are summarized in Table \ref{tb:results_meas}. We can see an improvement in terms of the positioning accuracy compared to a simple path-loss model but the benefit is not as large as expected from the synthetic results. An interesting aspect of the results is, that even though the results with neglected orientation are in good accordance with the theoretical results, we were not able to capture all additional effects with the antenna pattern correctly. One explanation for this discrepancy is that we use an antenna pattern based on the statistics of the measurement. We did not perform any calibration or measurements in an anechoic chamber previous to the measurement campaign. Also the assumption that all antenna patterns are the same does not have to be true due to varying parameters in the production chain and a strong influence of the metal shelves. Nevertheless, our results show that it is possible to achieve a position RMSE below one meter with measurements from low-power and low-bandwidth sensors in an indoor environment with many OLOS conditions. 

\subsection{Increased number of anchors}
In this section, we investigate the impact of doubling the amount of anchors on the positioning performance, which results in one anchor at the lowest and one anchor at the highest level at each corner of a shelf (see Fig.~\ref{fig:synthScen}). We perform the same calculations as in the previous sections for synthetic and measured data. The comparison for both cases is given in Table \ref{tb:results_synth} and Table \ref{tb:results_meas}, respectively. We can see that doubling the number of anchors has only a small impact on the positioning accuracy. The number of cooperative measurements is ${N_C = |\mathcal{C}| \times (|\mathcal{C}|-1)/2}$ and the number of anchor measurements ${N_A = |\mathcal{A}| \times |\mathcal{C}|}$. In this scenario ${N_C \gg N_A}$ which means that most of the information about the position of the agents is gained due to cooperation. 
\subsection{Parameter estimation with subset of nodes}
\label{subsec:nodeSubset}
\begin{figure}[t]
\centering
\includegraphics[scale=1]{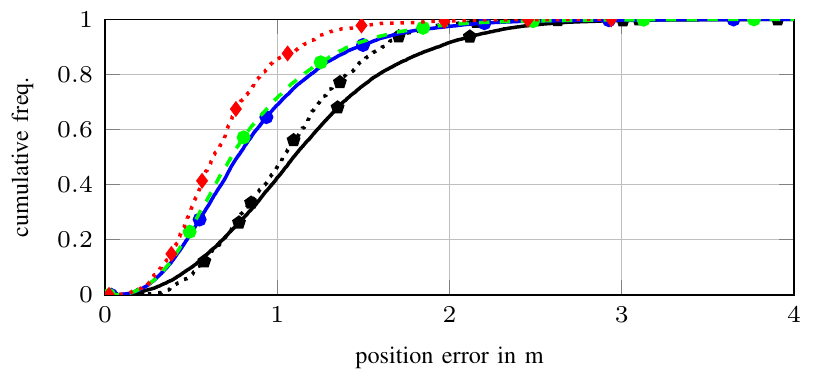}\hfill
\includegraphics[scale=1]{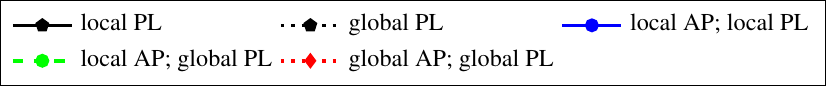}\hfill\\
\caption{This figure shows the CF of the positioning error with regard to local and global parameter estimation. The abbreviations 'PL' and 'AP' stand for path-loss and antenna pattern respectively.}
	\label{fig:CDF_Window}
\end{figure}

In this section, we show how estimating the model parameters based on a subset of nodes influences the overall positioning performance. For that purpose, we split the scenario in 15 equally spaced areas in which the model parameters are estimated based on \eqref{eq:argmax}. Each area has a length in the x-direction of 3.5~m and a width in the y-direction of 1~m. The center of an area is in the middle of a corridor (see Fig.~\ref{fig:synthScen}). For each estimated parameter set, we use 4000 particles to evaluate the whole scenario. In addition, we investigate the influence on the positioning performance if only the path-loss is estimated in each area and if the parameters of the antenna pattern in each area are estimated based on the global path-loss. The global parameters are determined with respect to all nodes in the scenario. The results in terms of the cumulative frequency of the positioning error can be seen in Fig.~\ref{fig:CDF_Window}. Note that the largest outlier for the results with local path-loss is at 12~m. The results show that using only a locally estimated path-loss and neglecting the antenna pattern has the worst performance. Using the global estimate for the path-loss and local estimates for the antenna pattern has no real performance gain compared to using only locally estimated parameters. The global parameter set achieves the highest positioning accuracy but using local estimates of the parameters leads to comparable results, which indicates that it should be beneficial to develop an algorithm that jointly estimates position, orientation, and the model parameters.
\section{Conclusion}
\label{sec:Conclusion}
In this paper, we propose two methods for jointly estimating position and orientation of a cooperative network based on factor graphs. The first method uses a continuous representation of the orientation state whereas the second one uses a discrete representation. We employ RSS-based ranging with an RSS model that depends also on the orientation of the nodes. This directivity is modelled via an antenna pattern. We validate our proposed methods with simulations and real measurements for more than 900 agents in an indoor environment and show that the positioning performance can be increased significantly. We achieve a position RMSE below 0.80~m on an area of 112~m$^2$ and a height of 1~m. The large number of agents leads to a large number of measurement per agent with less influence of outliers, since it is possible to represent the statistics of the measurements more precisely. We also investigate the impact of estimating the model parameters in subregions of the scenario on the overall positioning performance. We see that global and local parameter estimation achieve comparable results. Our current research focuses on including the estimation of the model parameters in the belief propagation algorithm. Possible directions of future research could deal with the question how to reduce the number of measurements to minimize the energy consumption.

\section*{Acknowledgment}
\label{sec:Acknowledgement}
We want to thank SES-imagotag GmbH for the support of the project and measurement campaign, our colleagues from TU Wien and TU Graz for their help with the preparation and execution of the measurement campaign and the staff of the library for their support and patience.

\bibliographystyle{IEEEtran}
\bibliography{IEEEabrv,bibfile}

\end{document}